\pdfoutput=1
\RequirePackage{etoolbox}
\csdef{input@path}{%
 {img/}%
}%
\csgdef{bibdir}{bib/}

\documentclass[ba,noinfoline]{imsart}
\pubyear{0000}
\volume{00}
\issue{0}
\doi{0000}
\firstpage{1}
\lastpage{1}

\usepackage{amsmath} 
\usepackage{natbib} 
\usepackage[colorlinks,citecolor=blue,urlcolor=blue,filecolor=blue,backref=page]{hyperref} 
\usepackage{caption} 
\usepackage{subfigure} 
\usepackage{enumitem}
\usepackage[toc]{appendix}
\usepackage{mathtools} 
\usepackage[]{algorithm2e}
\usepackage[noabbrev]{cleveref}
\usepackage{sidecap}
\sidecaptionvpos{figure}{c}
\usepackage{booktabs} 
\newcommand{\overbar}[1]{\mkern1.5mu\overline{\mkern-1.5mu#1\mkern-1.5mu}\mkern1.5mu} 
\newcommand{\programscript}[1]{{\sc #1}}              
\newcommand{\PolyChord}{\programscript{PolyChord}}     
\newcommand{\MultiNest}{\programscript{MultiNest}}     
\newcommand{\ecs}{colour scale shows the fraction of the cumulative probability distribution lying between some region and the median.} 
\newcommand{\Explaincs}{The \ecs}

\newcommand{\eb}{in brackets show the error on the final digit.}
\newcommand{\Explainbrackets}{Numbers \eb}
\newcommand{\explainbrackets}{numbers \eb}
\newcommand{\like}{\ensuremath{\mathcal{L}}}
\renewcommand{\d}[1]{\ensuremath{\operatorname{d}\!{#1}}}
\newcommand{\Z}{\ensuremath{\mathcal{Z}}}
\newcommand{\e}{\mathrm{e}}
\newcommand{\logx}{\ensuremath{\log X}}
\newcommand{\logZ}{\ensuremath{\log \Z}}

\newcommand{\bt}{\mathbf{t}}
\newcommand{\thetaone}{\theta_{\hat{1}}}
\newcommand{\ndead}{n_\mathrm{dead}}
\newcommand{\po}{\overbar{\theta_{\hat{1}}}}
\newcommand{\conf}[2]{\ensuremath{\mathrm{C.I.}_{#2 \%}(#1)}}
\newcommand{\poconf}[1]{\conf{\po}{#1}}
\newcommand{\E}{\mathrm{E}}
\newcommand{\Var}{\mathrm{Var}}
\newcommand{\std}{\ensuremath{\mathrm{St.Dev.}}}
\newcommand{\wit}{w_i (\mathbf{t})}
\newcommand{\expect}[1]{\ensuremath{\E \left[ #1 \right]}}
\newcommand{\Cov}{\mathrm{Cov}}
\newcommand{\bx}{\mathbf{x}}
\newcommand{\ftilde}{\tilde{f}}
\newcommand{\vart}{\overbar{\theta_{\hat{1}}^2}} 
\newcommand{\stdct}[1]{\ensuremath{\std\!\left[\left\{ #1 \right\}\right]}}

\begin{document}

\begin{frontmatter}

\title{Sampling errors in nested sampling parameter estimation}

\begin{aug}
\author{\fnms{Edward} \snm{Higson}\thanksref{t1,addr1,addr2}\ead[label=e1]{e.higson@mrao.cam.ac.uk}},
\author{\fnms{Will} \snm{Handley}\thanksref{t2,addr1,addr2}\ead[label=e2]{wh260@mrao.cam.ac.uk}},
\author{\fnms{Mike} \snm{Hobson}\thanksref{t3,addr1}\ead[label=e3]{mph@mrao.cam.ac.uk}}
\and
\author{\fnms{Anthony} \snm{Lasenby}\thanksref{t4,addr1,addr2}\ead[label=e4]{a.n.lasenby@mrao.cam.ac.uk}}
\runauthor{E. Higson et al.}
\thankstext{t1}{\printead{e1}}
\thankstext{t2}{\printead{e2}}
\thankstext{t3}{\printead{e3}}
\thankstext{t4}{\printead{e4}}
\address[addr1]{Astrophysics Group, Battcock Centre, Cavendish Laboratory, JJ Thomson Avenue, Cambridge CB3 0HE, UK}
\address[addr2]{Kavli Institute for Cosmology, Madingley Road, Cambridge, CB3 0HA, UK}
\end{aug}

\begin{abstract}
Sampling errors in nested sampling parameter estimation differ from those in Bayesian evidence calculation, but have been little studied in the literature.
This paper provides the first explanation of the two main sources of sampling errors in nested sampling parameter estimation, and presents a new diagrammatic representation for the process.
We find no current method can accurately measure the parameter estimation errors of a single nested sampling run, and propose a method for doing so using a new algorithm for dividing nested sampling runs.
We empirically verify our conclusions and the accuracy of our new method.
\end{abstract}

\begin{keyword}
\kwd{nested sampling}
\kwd{parameter estimation}
\end{keyword}

\end{frontmatter}

\section{Introduction}

Nested sampling~\citep{Skilling2006} is a Monte Carlo method for Bayesian analysis which simultaneously calculates both Bayesian evidences and posterior samples.
The early development of the algorithm was focused on evidence calculation, which is computationally expensive using variants of standard Markov chain Monte Carlo (MCMC) sampling based on the Metropolis-Hastings algorithm~\citep{Mackay2003}.

Contemporary implementations such as \MultiNest{} \citep{Feroz2008,Feroz2009,Feroz2013} and \PolyChord{} \citep{Handley2015a,Handley2015b} are now also extensively used for parameter estimation from posterior samples~\citep[see for example][]{PlanckCollaboration2016}.
Nested sampling compares favourably to MCMC-based parameter estimation for degenerate, multi-modal likelihoods as it has no ``thermal'' transition probability and exponentially compresses the prior distribution to the posterior.
However, despite its increasing popularity, the sampling errors in nested sampling parameter estimation are poorly understood.

Correctly quantifying uncertainty is vital for identifying spurious results --- in particular we find sampling errors often significantly affect estimates of credible intervals on parameters.
Conversely, finding such errors are very small may imply an unnecessarily large amount of computational resource is being used for the calculation.
This paper has two goals: to provide an explanation of the sources of these errors and an empirical technique for estimating them.
One obvious method is to repeat the analysis some $n_{\mathrm{repeats}}$ times, although this increases the computational cost by a corresponding factor.
Interestingly, we find no current method can accurately estimate these errors on parameter estimates from a single analysis, and so we present a new method for doing this.
Our approach uses a new algorithm for dividing a single nested sampling run into multiple valid nested sampling runs; these can then be recombined in different combinations using resampling techniques such as the bootstrap.
We test our results and new method empirically.

The paper begins with background on sampling errors in parameter estimation from posterior samples, then describes the nested sampling algorithm and how it is currently used for parameter estimation in Section~\ref{sec:nested_sampling_algorithm}.
We explain the two main sources of sampling errors in nested sampling parameter estimation in Section~\ref{sec:theory}, and present a new diagrammatic representation of the process (illustrated in~\Cref{sub:p1_gaussian,sub:p1_gaussian_3d,sub:p1_cauchy,sub:p1var_gaussian,sub:r_gaussian}).
Section~\ref{sec:multiple_runs} describes our new method for measuring sampling errors from a single nested sampling run, using our new algorithm for division of such runs.

We empirically test our method's accuracy in~\Cref{sec:numerical_tests} with the help of analytical cases in the manner described by~\citet{Keeton2011}.
Here one can obtain uncorrelated samples from the prior space within some likelihood contour using standard techniques, and we term the resulting procedure {\em perfect nested sampling}.
Our approach accurately quantifies uncertainties on parameter estimates from the stochasticity of the nested sampling algorithm, but software used for practical problems may produce additional errors from correlated samples within likelihood contours that are specific to a given implementation.
We discuss implementation-specific errors in~\Cref{sec:implementation_specific}, including testing sampling error estimates from our method for \PolyChord{} calculations.
Our method gives superior performance to the current approach and can be easily included in nested sampling software; we are currently working on incorporating it into future versions of \PolyChord{}.

\subsection*{Background: sampling errors in parameter estimation}

Sampling can be used to represent a posterior distribution $\mathcal{P}(\theta)$ via a set of weighted samples 
\begin{equation}
  \mathcal{S} = \{(\theta_s, p_s), s=1,\ldots, n_\mathrm{samp}\},
  \label{equ:samples}
\end{equation}
where each $\theta_s$ is drawn from the posterior distribution with probability proportional to $p_s\times\mathcal{P}(\theta_s)$, and $\sum_{s\in\mathcal{S}} p_s = 1$.
Likelihoods $\like$ are often computationally expensive functions, so the goal of parameter estimation is to sample the posterior distribution $\mathcal{P}(\theta)$ numerically with a limited number of likelihood calls.

Samples $\mathcal{S}$ may be used to compute numerical results. For example, the posterior expectation of a function of the parameters $f(\theta)$ can be estimated as
\begin{equation}
    \E[f(\theta)]
    =
    \int f(\theta)\mathcal{P}(\theta)\:d\theta
    \: \approx \:
    \sum_{s\in\mathcal{S}} p_s f(\theta_s).
    \label{equ:expec_sample_error}
\end{equation}
In this case the sampling error is the difference between $\sum_{s\in\mathcal{S}} p_s f(\theta_s)$ and the exact value of $\E[f(\theta)]$.
Often the posterior distributions of parameters $\theta$ are of interest, and are estimated numerically from the samples by dividing the parameter space into cells or via kernel density estimation.

There have been many works on approximating MCMC sampling errors, including investigation of quantiles and the amount of computation required to reach some level of accuracy --- see for example \citet{Doss2015,Flegal2008,Liu2016}.
In particular Sequential Monte Carlo samplers~\citep{DelMoral2006} have similarities with nested sampling, and their sampling errors are better understood.
For some related methods such as the Tootsie Pop algorithm~\citep{Huber2014} and accelerated simulated annealing~\citep{Bezakova2008} the error distribution is known exactly, although these techniques are less widely used.
This paper introduces empirically tested techniques for quantifying sampling errors from the nested sampling algorithm.

\section{The nested sampling algorithm}\label{sec:nested_sampling_algorithm}

Nested sampling~\citep{Skilling2006} is a numerical method computing Bayesian evidences
\begin{equation}
    \Z
    =
    \int \like(\theta) \pi (\theta)\d{} \theta
    \label{equ:Z_definition}
\end{equation}
and samples from the posterior distribution
\begin{equation}
    \mathcal{P}(\theta) = \frac{\like(\theta) \pi(\theta)}{\Z}
    \label{equ:parameter_estimation}
\end{equation}
given some likelihood $\like(\theta)$ and prior $\pi(\theta)$.

Initially $n$ points, termed {\em live points}, are sampled randomly from the prior.
At each iteration $i$, the live point with the lowest likelihood $\like_i$ is removed and replaced by a new live point sampled from the prior subject to the constraint that it has a likelihood higher than $\like_i$.
Iterating until some termination condition is met generates a list of discarded samples known as {\em dead points}, which are used to estimate the evidence and make posterior inferences%
\footnote{The remaining live points at termination can also be used if required, but termination conditions can be chosen such that this makes a negligible difference to calculation results.}%
.
We refer to the completed nested sampling process as a {\em run}.

\begin{figure}
    \centering
    \hspace*{0.09cm} 
    \includegraphics[width=\linewidth]{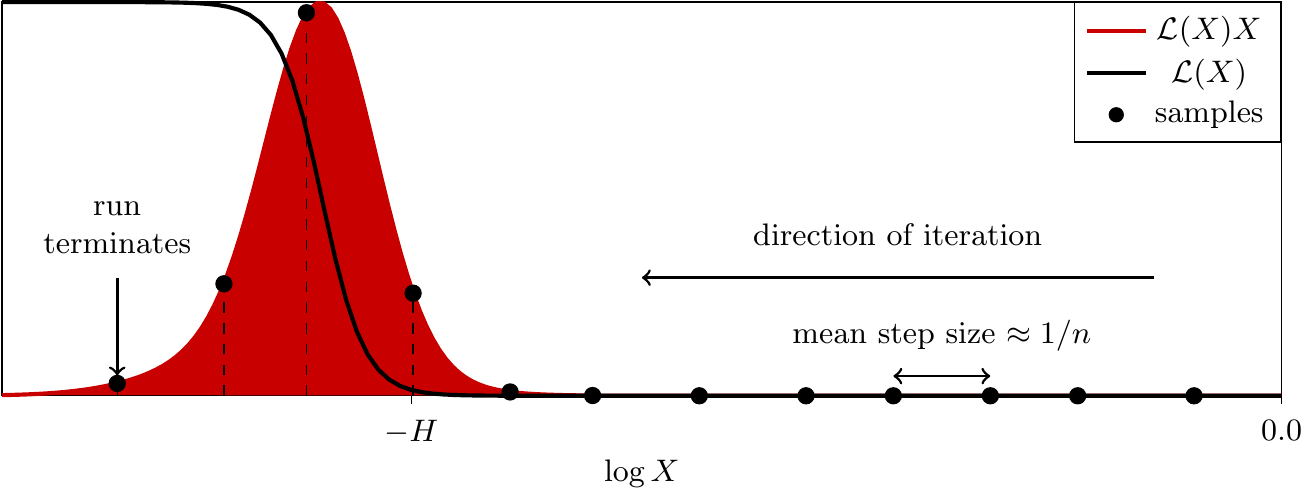}
    \caption{A schematic representation of nested sampling with a constant number of live points $n$.
The curve $\like(X)X$ shows the relative posterior mass, the bulk of which is contained in some small fraction $\exp(-H)$ of the prior and is only visible on a log scale in $X$.
The algorithm iterates inwards in $X$ exponentially with stochastic shrinkage ratios distributed according to~\eqref{equ:dist_t}.}\label{fig:ns_evidence}
\end{figure}

To compute the evidence, the many-dimensional integral~\eqref{equ:Z_definition} is reduced to a one-dimensional integral in terms of the fractional prior volume within an iso-likelihood contour.
We define the fraction of the prior $\theta$ with likelihood $\like(\theta)$ greater than some value $\like^\ast$ as $X(\like^\ast)$, where
\begin{equation}
    X(\like^\ast)\equiv \int_{\like(\theta)>\like^\ast} \pi(\theta)\d \theta,
\end{equation}
and $X \in [0,1]$. Provided the inverse $\like(X)\equiv X^{-1}(\like)$ exists\footnote{A sufficient condition for $\like(X)\equiv X^{-1}(\like)$ to exist is for $\like$ to be continuous and $\pi$ to have a connected support.
See \citet{Chopin2010} and \citet[][Appendix C]{Feroz2013} for a more detailed measure-theoretic discussion.}, the evidence~\eqref{equ:Z_definition} can be expressed as
\begin{equation}
    \Z=\int_0^1 \like(X) \d X.
    \label{equ:Z(X)}
\end{equation}
Given a set of dead points with likelihoods $\like_i$, the corresponding prior volumes $X_i$ are unknown but are modelled statistically as $X_i = t_i X_{i-1}$, where $X_0 = 1$ and each shrinkage ratio $t_i$ is independently distributed as the largest of $n$ random variables from the interval $[0,1]$ \citep{Skilling2006}.
Hence:
\begin{equation}
    P(t_i)            = n t_i^{n-1}, \qquad
    \E[\log t_i ]        = -\frac{1}{n},   \qquad
    \mathrm{Var}[\log t_i ]    = \frac{1}{n^2},
    \label{equ:dist_t}
\end{equation}
and the algorithm samples within an exponentially shrinking part of the prior.
This exponential shrinkage is shown schematically in~\Cref{fig:ns_evidence}.

\subsection{Evidence estimation}\label{sec:sampling_evidence_error}

Nested sampling therefore allows one to approximate the evidence~\eqref{equ:Z(X)} via a quadrature sum over the dead points
\begin{equation}
    \Z(\bt) \approx \sum_{i \in \mathrm{dead}} \like_i w_i(\mathbf{t}),
    \label{equ:ztot}
\end{equation}
where $\bt=\{t_1,t_2,\dots,t_{\ndead}\}$ are the unknown set of shrinkage ratios for the $\ndead$ iterations of the nested sampling process, and each $t_i$ is an independent random variable drawn from distribution~\eqref{equ:dist_t}.
The shrinkage ratios define the prior volumes via $X_i(\bt) = \prod^i_{k=0} t_k$, and the $w_i$ are appropriately chosen quadrature weights roughly corresponding to the volume of the ``prior shell'' to which a given dead point belongs. For example, using the trapezium rule: $w_i(\mathbf{t})=\frac{1}{2}(X_{i-1}(\mathbf{t})-X_{i+1}(\mathbf{t}))$.

Given that the shrinkage ratios $\bt$ are a priori unknown, we may quantify our knowledge of $\Z$ by simulating sets of $\bt$ according to~\eqref{equ:dist_t}, and working with the distribution of the resulting set of evidences ${\{\Z\}}_{\bt}$ from~\eqref{equ:ztot}~\citep{Skilling2006}. Typically one then computes and reports a mean value and error for $\logZ$ from this distribution.

Several alternative methods for calculating evidence inferences are reported in the literature. 
\citet{Skilling2006} also proposes an error calculation based on relative entropy, which demonstrates that the uncertainty of $\logZ$ is dominated by the Poisson variability in the number of steps required to reach the bulk of the posterior mass.
\citet{Keeton2011} uses distribution moments and running totals which are updated with each nested sampling step.
This method has been extended by~\citet{Handley2015b} to allow the splitting of multi-modal likelihoods into different clusters and the treatment of variable numbers of live points.
For a more detailed discussion of the convergence properties of nested sampling evidences, see~\citet{Chopin2010}.

Thus, the dominant sampling error in the evidence estimate~\eqref{equ:ztot} from perfect nested sampling is from statistical variation in the unknown volumes of the prior ``shells'' $\wit$ that each point represents.
The error from approximating the integral for $\Z$ with a sum can be safely neglected unless $n$ is very small\footnote{The trapezium rule error is $\mathcal{O}(1/n^2)$, and if required other methods such as Simpson integration could be used.}~\citep{Skilling2006}.
There is also some error from terminating the algorithm and truncating the sum, but this is can be made negligible with appropriate termination conditions.

\subsection{Parameter estimation}\label{sec:sampling_parameter_error} 

One may also perform posterior inference from nested sampling by using the dead points to construct a set of posterior samples with weights proportional to their share of the posterior mass~\citep{Skilling2006}:
\begin{equation}
    p_i(\mathbf{t})=\frac{w_i(\mathbf{t})\like_i}{\sum_i w_i(\mathbf{t})\like_i}=\frac{w_i(\mathbf{t})\like_i}{\Z(\mathbf{t})}.
    \label{equ:posterior_weight}
\end{equation}
As before, $\mathbf{t}$ is the set of prior shrinkage ratios and in the trapezium rule case $w_i(\mathbf{t}) = \frac{1}{2}(X_{i-1}(\mathbf{t})-X_{i+1}(\mathbf{t}))$.

The weights defined by~\eqref{equ:posterior_weight} present a departure from traditional sampling approaches in that the $w_i(\mathbf{t})$ are random variables, with their stochasticity determined by~\eqref{equ:dist_t}.
When computing expectations~\eqref{equ:expec_sample_error} there is now an additional error associated with our lack of knowledge of the precise values $p_i(\mathbf{t})$.
Nested sampling software packages such as \MultiNest{} and \PolyChord{} produce posterior files containing only the expected values
\begin{equation}
    \E[p_i(\bt)] = \frac{\e^{-i/n}\mathcal{L}_i}{\sum_{j} \e^{-j/n}\mathcal{L}_j}.
    \label{equ:param_exp}
\end{equation}

To account for the stochasticity in the weights $p_i$,~\citet{Skilling2006} suggests simulating the prior volume shrinkage ratios $\mathbf{t}$ in the same manner as for evidence estimation, and using these simulations to calculate a set of values for estimators such as~\eqref{equ:expec_sample_error}.
The sampling error should then be estimated from the variation within this sample; we term this the {\em simulated weights method}.
We believe this procedure is the only estimate of sampling errors in parameter estimation from a single nested sampling run proposed in the literature.
However it is in general an {\em underestimate}, as can be seen in the numerical tests in~\Cref{sec:numerical_tests}.
Section~\ref{sec:analytical_sqrt_2} of the supplementary material discusses this underestimation of errors in detail.

We now describe why the simulated weights method does not capture all sources of sampling errors, and in Section~\ref{sec:multiple_runs} we propose a new method for correctly computing these errors.

\section{Sources of sampling errors in nested sampling parameter estimation}\label{sec:theory}

\begin{SCfigure}
\includegraphics[width=0.65\linewidth]{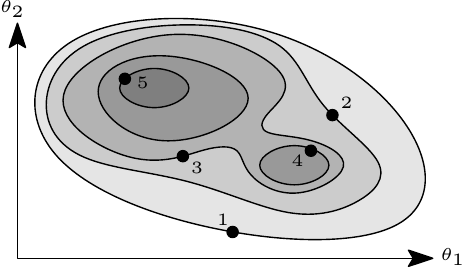}
\caption{%
Nested sampling dead points and iso-likelihood contours for a two-dimensional multi-modal likelihood $\like(\theta)$; darker shading shows higher likelihoods.
Iso-likelihood contours can pass through a wide range of different parameter values.
}\label{fig:will_contours}
\end{SCfigure}

In order to understand why the simulated weights method underestimates sampling errors, we require a result from~\citet{Chopin2010}.
They show that the expectation integral~\eqref{equ:expec_sample_error} may be re-phrased in terms of the prior volume $X$ via:
\begin{equation}
    \E[f(\theta)]=
    \int f(\theta)\mathcal{P}(\theta)\:d\theta 
    =\int f(\theta)\frac{\like(\theta)\pi(\theta)}{\Z}\d\theta
    =\frac{1}{\Z}\int \ftilde(X)\like(X)\d X,
    \label{equ:Z(X)_integral}
\end{equation}
where $\ftilde(X)$ is the prior expectation of $f(\theta)$ on some iso-likelihood contour $\like(\theta)=\like(X)$,
\begin{equation}
    \ftilde(X)\equiv\E^{\pi}[f(\theta)|\like(\theta)=\like(X)].
    \label{equ:f(X)}
\end{equation}
The simulated weights approach amounts to discretising the integral~\eqref{equ:Z(X)_integral} as
\begin{equation}
    \frac{1}{\Z}\int \ftilde(X)\like(X)\d X
    \approx
    \frac{1}{\Z}\sum_i \ftilde(X_i)\: \like_i\:  \frac{1}{2}(X_{i-1} - X_{i+1}),
    \label{equ:discretise}
\end{equation}
and, most importantly, further requiring that we may use $f(\theta_i)$ as a proxy for $\ftilde(X_i)$ at each point $X_i$,
In some special cases $f(\theta_i)=\ftilde(X_i)$ for all $\theta$ and this approach is valid, for example when $f(\theta_i)=\ftilde(X_i) \propto -\log\like_i$ (entropy computation), but in general it is not.
This can cause significant inaccuracies as iso-likelihood contours often span wide ranges of different parameter values, as illustrated in~\Cref{fig:will_contours}.

To summarise, the dominant sampling errors in estimating some parameter or function of parameters from perfect nested sampling typically come from two sources:
\begin{enumerate}[label= (\roman*)]
    \item approximating the unknown prior volumes $\wit$ with their expectation $\E[\wit]$ using~\eqref{equ:dist_t};\label{enu:w_error}
    \item approximating the mean value of a function of parameters over an entire iso-likelihood contour $\ftilde(X_i)$ with its value at a single point $f(\theta_i)$.\label{enu:sample_error}
\end{enumerate}
Errors from~\ref{enu:w_error} are also present in evidence calculation; in the parameter estimation case they are typically smaller as results depend only on the {\em relative\/} weights of the samples.
In contrast~\ref{enu:sample_error} is only present in parameter estimation, where it is typically a significant or dominant source of sampling errors.
The relative contributions of~\ref{enu:w_error} and~\ref{enu:sample_error} are empirically tested in Section~\ref{sec:relative_errors} of the supplementary material, where they are calculated for analytical cases by using exact values for weights $\wit$ and by replacing $f(\theta_i)$ with $\ftilde(X_i)$.
The simulated weights method underestimates sampling errors in parameter estimation as it ignores errors from~\ref{enu:sample_error}.

We now introduce a new diagrammatic representation of nested sampling parameter estimation to illustrate the two different sources of sampling errors.

\subsection{Diagrammatic representation}\label{sec:diagram}

Nested sampling transforms evidence calculations of any dimension into a 1-dimensional problem\footnote{\label{foo:equivalence}For practical nested sampling problems implementation-specific errors can differ for two likelihoods with the same $\like(X)$.
For example if one likelihood has a much higher dimension and a much larger number of modes than the other it may have larger errors from the implementation software.}
in $\like(X)$ which can be entirely represented on a diagram like Figure~\ref{fig:ns_evidence}.
An analogous diagram for parameter estimation must also illustrate sampling a single point $f(\theta_i)$ on each iso-likelihood contour $\like(\theta)=\like(X_i)$ from the distribution $P(f(\theta)|X_i)$.

We propose a generalisation of Figure~\ref{fig:ns_evidence} for visualising parameter estimation problems, and present it in~\Cref{sub:p1_gaussian,sub:p1_gaussian_3d,sub:p1_cauchy,sub:p1var_gaussian,sub:r_gaussian}.
The top panel in each figure is similar to Figure~\ref{fig:ns_evidence} and shows the relative posterior mass $\like(X)X$ at each value of $\logx$.
The lower central panel shows the probability distribution $P(f(\theta)|X)$ and its mean $\ftilde(X)$.
The posterior distribution is shown on the left --- this is equal to the distributions $P(f(\theta)|X)$ (the lower central panel) marginalised over $X$ in proportion to the posterior weight at each $X$ (the top panel).

\newcommand{\figtrim}{0.04in} 
\begin{figure}
\centering
\subfigure[%
$f(\theta)=\theta_{\hat{1}}$ with a 5-dimensional Gaussian likelihood~\eqref{equ:gaussian} and a Gaussian prior~\eqref{equ:gaussian_prior}.
]{%
\includegraphics[trim={0 0 0 \figtrim},clip,width=\linewidth]{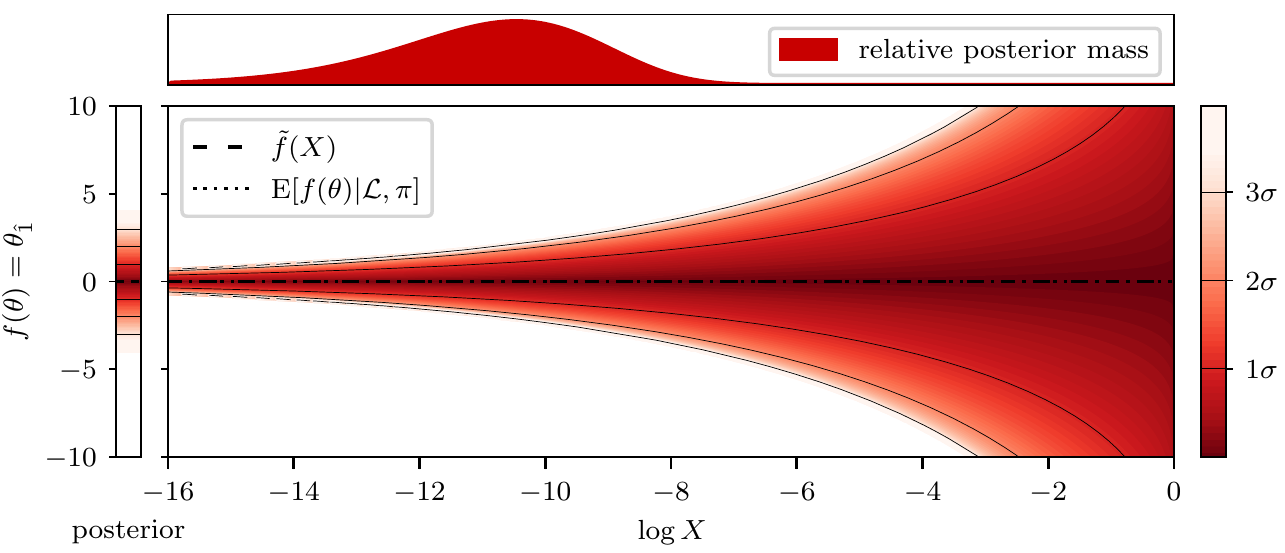}%
\label{sub:p1_gaussian}}\\
\subfigure[%
$f(\theta)=\theta_{\hat{1}}$ with a 5-dimensional Cauchy likelihood~\eqref{equ:cauchy} and a Gaussian prior~\eqref{equ:gaussian_prior}.
]{%
\includegraphics[trim={0 0 0 \figtrim},clip,width=\linewidth]{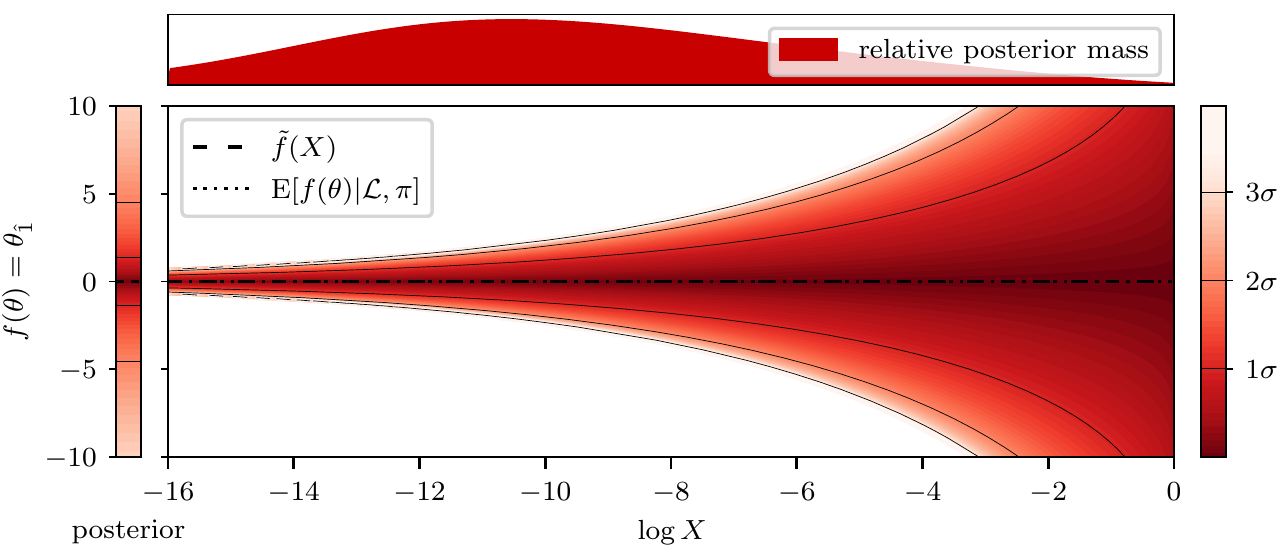}%
\label{sub:p1_cauchy}}\\
\centering
\subfigure[%
$f(\theta)=\theta_{\hat{1}}$ with a 3-dimensional Gaussian likelihood~\eqref{equ:gaussian} and a Gaussian prior~\eqref{equ:gaussian_prior}.
]{%
\includegraphics[trim={0 0 0 \figtrim},clip,width=\linewidth]{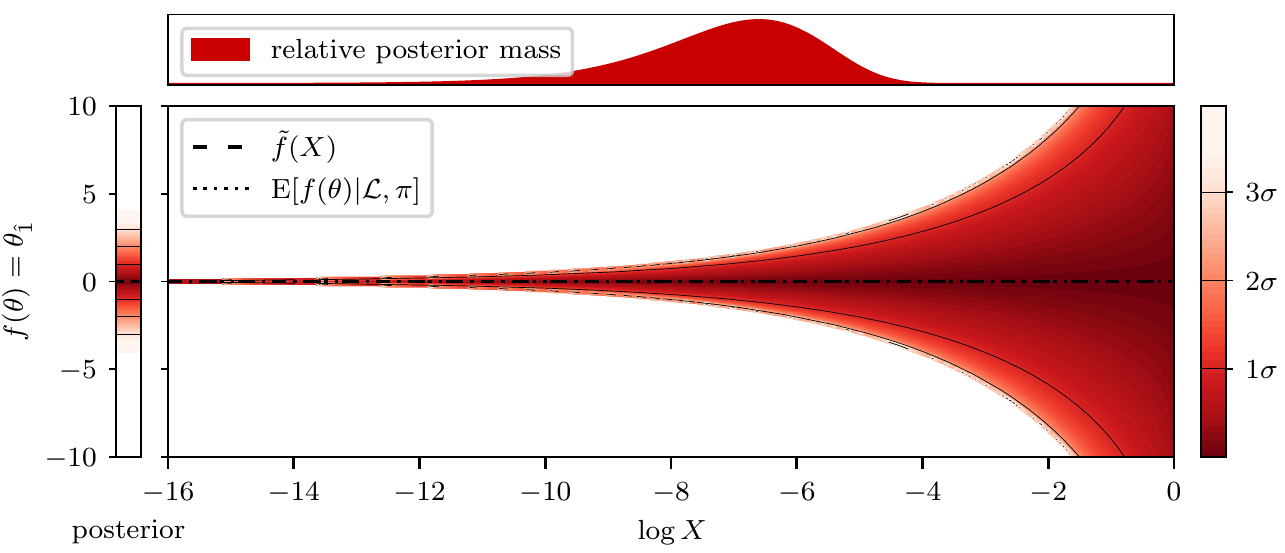}%
\label{sub:p1_gaussian_3d}}\\
\end{figure}
\begin{figure}
\subfigure[%
$f(\theta)={\thetaone}^2$ with a 5-dimensional Gaussian likelihood~\eqref{equ:gaussian} and a Gaussian prior~\eqref{equ:gaussian_prior}.
]{%
\includegraphics[width=\linewidth]{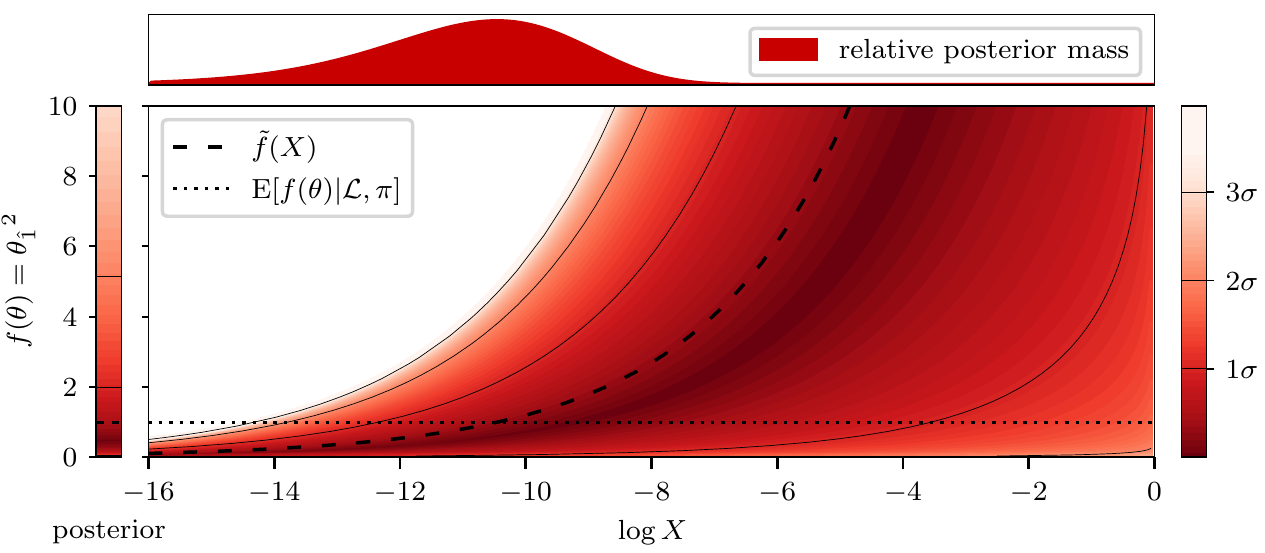}%
\label{sub:p1var_gaussian}}\\
\subfigure[%
$f(\theta)=|\theta|$ (i.e.\ the radial distance from the likelihood's maximum) with a 5-dimensional Gaussian likelihood~\eqref{equ:gaussian} and a Gaussian prior~\eqref{equ:gaussian_prior}.
In this case $f(\theta_i)=\ftilde(X_i)$ for all $\theta$ and sampling errors are only from uncertainty in prior volume shrinkages and the trapezium rule approximation.
]{%
\includegraphics[trim={0 0 0 \figtrim},clip,width=\linewidth]{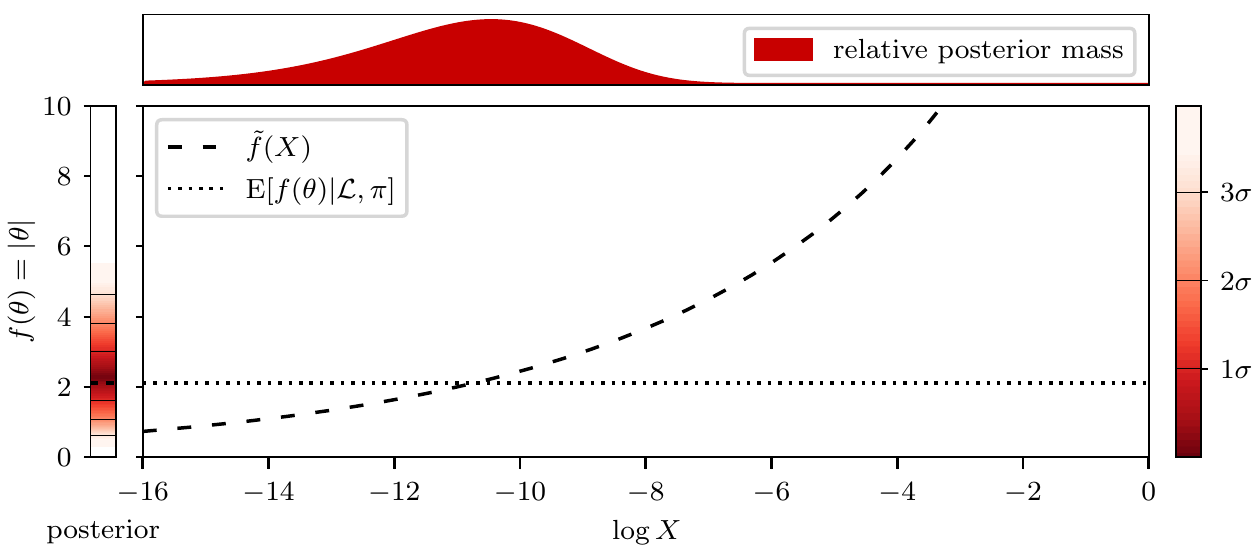}%
\label{sub:r_gaussian}}\\
\caption{%
Nested sampling parameter estimation diagrams:
in each case the top panel shows the relative posterior mass at each value of $\logx$ ($\propto \like(X) X$). The lower central panel shows the distribution $P(f(\theta)|X)$ of values $f(\theta)$ on each iso-likelihood contour $\like(\theta)=\like(X)$; the dashed line shows the expectation of this distribution which we defined in~\eqref{equ:f(X)} as $\ftilde(X)$. The left panel shows the posterior distribution of $f(\theta)$, with the dotted line showing its posterior expectation.
\Explaincs{}
}\label{fig:ns_param}
\end{figure}

For these example plots we use $d$-dimensional spherical unit Gaussian likelihoods
\begin{equation}\label{equ:gaussian}
    \like(\theta) = {(2 \pi)}^{-d/2} \e^{-{|\theta|}^2 / 2}
\end{equation}
and $d$-dimensional spherical unit Cauchy likelihoods
\begin{equation}\label{equ:cauchy}
    \like(\theta)=\frac{\Gamma(\frac{1+d}{2})}{\pi^{(d+1)/2}}{\left(1+{|\theta|}^2\right)}^{-(\frac{d+1}{2})},
\end{equation}
with $d$-dimensional co-centred spherical Gaussian priors
\begin{equation}\label{equ:gaussian_prior}
    \pi(\theta) = {(2 \pi \sigma_\pi^2)}^{-d/2} \e^{-{|\theta|}^2 / 2 \sigma_\pi^2}, \qquad \sigma_\pi = 10.
\end{equation}
We denote the first component of the $\theta$ vector as $\theta_{\hat{1}}$, although by symmetry the results will be the same for any component.
$\po$ and $\vart$ are the first and second moments of the posterior distribution of $\theta_{\hat{1}}$.

The form of the distribution $P(f(\theta)|X)$ as $X$ varies depends on the likelihood only through the shape of the iso-likelihood contours $\like(\theta)=\like(X)$.
Therefore the lower central panel of the diagrams for some $f(\theta)$ is the same for any likelihoods with the same contours --- this can be seen in~\Cref{sub:p1_gaussian,sub:p1_cauchy}, where the differences in the posterior (left panel) are due only to the different posterior weights in $\logx$ (top panel).

These diagrams can be constructed for any nested sampling calculation using the posterior samples and kernel density estimation; this could provide insight into the nature of the calculation and the relative contributions from the different sources of sampling errors.

\subsection{Transforming a parameter estimation problem into 2 dimensions}\label{sec:2d_representation}

As illustrated by our diagrams, nested sampling parameter estimation is fundamentally a 2-dimensional problem in $\like(X)$ and $P(f(\theta)|X)$.
In fact a parameter estimation calculation for some $f(\theta)$ given $\like(\theta)$ is equivalent to a 2-dimensional problem for $f^\ast(\theta^\ast)$ given $\like^\ast(\theta^\ast)$ when
\begin{align}
    \like^\ast(\theta^\ast) &= \like(X),\label{equ:X_isomorph_condition} \\
    P(f^\ast(\theta^\ast)|X) &= P(f(\theta)|X), \label{equ:f_isomorph_condition}
\end{align}
for all $X$.
Any transformation satisfying~\eqref{equ:X_isomorph_condition} and~\eqref{equ:f_isomorph_condition} will leave our proposed diagram for the calculation unchanged.
Parameter estimation can also be represented as a 1-dimensional problem in $\like^\ast(\theta^\ast)=\like(X)$ combined with a univariate stochastic process for each dead point $i$ with the distribution $P(f(\theta)|X_i)$.

One way to express a general nested sampling calculation in 2 dimensions is to map it onto the unit square $\theta^\ast=(X,Y)$ with uniform priors $X,Y \in [0,1]$ and a likelihood $\like^\ast(\theta^\ast) = \like(X)$ which is independent of $Y$ and satisfies~\eqref{equ:X_isomorph_condition}.
In this case $X$ is as before the remaining fractional prior volume and $Y$ parameterises each iso-likelihood contour.
Using inverse transform sampling, for a general $f(\theta)$ a corresponding $f^\ast(\theta^\ast)$ satisfying~\eqref{equ:f_isomorph_condition} is
\begin{equation}
    f^\ast(\theta^\ast) = f^\ast(X,Y) = F^{-1}(Y|X),
\end{equation}
where $F^{-1}(Y|X)$ is the inverse of the cumulative distribution
\begin{equation}
    F(Y|X) = \int^Y_{-\infty} P(f(\theta)=h|X) \d h.
\end{equation}

As an example let us consider $d$-dimensional spherically symmetric likelihoods such as~\eqref{equ:gaussian} or~\eqref{equ:cauchy} with co-centred spherically symmetric priors such as~\eqref{equ:gaussian_prior}.
Then $X(\theta)$ is a function only of the radial distance from the centre $|\theta|$, and the iso-likelihood contours $\like(\theta)=\like(X)$ are hyperspherical shells of some radius $|\theta_X |$.
The probability distribution of a single parameter $\thetaone$ (a single component of $\theta$) on such an iso-likelihood contour is then
\begin{equation}
    P(\theta_{\hat{1}}|X)
    =
    \begin{cases}
    \frac{\Gamma\!\left(\frac{d}{2}\right)}{|\theta_X| \, \Gamma(\frac{1}{2}) \, \Gamma\!\left(\frac{d-1}{2}\right)}
    {\left(1 - \frac{\theta_{\hat{1}}^2}{{|\theta_X|}^2} \right)}^{\frac{d-3}{2}}
    &
    \mathrm{if}\,-|\theta_X|<\theta_{\hat{1}}<|\theta_X|,
    \\
    0
    &
    \mathrm{otherwise}.
    \end{cases}
\label{equ:ndim_t1_dist}
\end{equation}
$\theta_{\hat{1}}$ can be sampled directly or used to calculate the inverse cumulative distribution which together with knowledge of the function $\like(X)$ allows the parameter estimation of a $d$-dimensional Gaussian to be transformed into a 2-dimensional problem on the unit square.

Samples from~\eqref{equ:ndim_t1_dist} can be generated efficiently using the symmetry around $\theta_{\hat{1}}=0$ and the change of variables $\Theta=\theta_{\hat{1}}^2 / {|\theta_X|}^2$ to give a Beta distribution
	\begin{align}
        P(\Theta|X)
        &=
        \begin{cases}
        \frac{\Gamma\!\left(\frac{d}{2}\right)}{\Gamma\!\left(\frac{1}{2}\right) \: \Gamma\!\left(\frac{d-1}{2}\right)} {\Theta}^{- \frac{1}{2}} {\left(1 - \Theta \right)}^{\frac{d-3}{2}}
        &
        \mathrm{if}\,0<\Theta<1,
        \\
        0
        & \mathrm{otherwise},
    \end{cases} \\
        \Theta & \sim \mathrm{Beta}\left(\frac{1}{2},\frac{d-1}{2}\right).
\end{align}
This technique is used for the numerical tests in~\Cref{sec:numerical_tests}, and allows the efficient sampling of high dimensional spherically symmetric distributions where only a few parameters are of interest without generating all the remaining uninteresting parameters.

\section{Estimating sampling errors in nested sampling parameter estimation}\label{sec:multiple_runs}

Following the discussion of sources of sampling errors in~\Cref{sec:theory}, we seek a method for correctly calculating parameter estimation sampling errors from a single nested sampling run.
As no additional samples $\theta_i$ are available, a natural starting point is to utilise resampling techniques such as the jackknife~\citep{Tukey1958}, bootstrap~\citep{Efron1979} and Bayesian bootstrap~\citep{Rubin1981}, which estimate the uncertainty on inferences from a set of samples by calculating the variation when samples are re-weighted.

However, as described in Section~\ref{sec:sampling_parameter_error}, the uncertainty in nested sampling weights $\wit$ produces additional sampling errors which are unique to the nested sampling process.
These are not accounted for by na\"{\i}vely applying jackknives and bootstraps to posterior samples produced by nested sampling, and these approaches fail when tested numerically.
We instead require a method for dividing runs in a manner that preserves the statistical properties of nested sampling. No such method exists in the literature, so we present one in the remainder of this section.

\subsection{Dividing runs into threads}\label{sec:divide}
\citet{Skilling2006} describes how several nested sampling runs $r=1,2,\dots$ with $n^{(r)}$ live points may be combined simply by merging the dead points and sorting by likelihood value.
The combined sequence of dead points is equivalent to a single nested sampling run with $n=\sum_r n^{(r)}$ live points.

In fact, as we show now, the reverse procedure is also possible.
A nested sampling run with $n$ points can be unwoven into a set of $n$ valid nested sampling runs, each with $n^{(r)}=1$.
We term these single live point runs {\em threads}. 
During nested sampling, each dead point $i$ is replaced by a new point sampled uniformly within its iso-likelihood contour $\like(\theta)=\like_i$.
Starting from each initial live point that is generated, one may follow this sequence of replacements down the set of dead points. This sub-sequence of dead points is in fact a nested sampling run with $n=1$.  More formally:

\begin{algorithm}[H]\SetAlgoLined{}
    \KwResult{$n$ threads.}
    \KwData{Dead points and the iterations at which they were sampled for a nested sampling run with $n$ live points.}
    Rank dead points by likelihood in ascending order\;
    \While{$i \in n$}{make a new stack $i$\;
        select one of the initial points sampled at the start of the run\;
        move the point out to the stack $i$\;
        \While{iteration $<$ final iteration}{select point sampled at the iteration where previous point was replaced (``died'')\;
            move the point to the stack $i$\;}
    }
    \caption{Splitting a nested sampling run into threads.}%
\label{alg:splitting}
\end{algorithm}

A few points are worthy of note:
\begin{enumerate}
    \item splitting a run by randomly selecting some fraction of the dead points will not produce threads (i.e.\ single point nested sampling runs);
    \item one may split a given nested sampling run into separate runs with $n^{(r)}\ne1$ by first separating into threads, and then recombining threads as desired;
    \item the algorithm can be easily adapted for varying numbers of live points by permitting it to select multiple points on contours where $n$ increases. This can result in constituent threads stopping or dividing into multiple threads part of the way through the run;
    \item typically there is only one point which was sampled uniformly from the prior volume within each dead point $i$'s iso-likelihood contour $\like(\theta)=\like_i$ --- the point which replaced $i$. A sufficient condition for a nested sampling run to only have one unique division into threads is that $\like(X)$ is an injective function;
    \item in order for the threads to be true nested sampling runs, care must be taken with the termination conditions conditions used. See Section~\ref{sec:termination} of the supplementary material for a full discussion.
\end{enumerate}

Given that threads represent independent nested sampling runs, one may apply standard resampling techniques to the set of threads and approximate the entire sampling error distribution without making assumptions about its form.
This works as the $\log X_i$ values of the dead points $i$ from some run with $n$ live points form a Poisson process with rate $n$, meaning the $\log X_j$ values of the dead points $j$ of a single thread are a Poisson process of rate 1.
For typical problems with computationally expensive likelihoods the computational cost of even a large number of resampling replications is negligible.

Having introduced a framework for applying resampling to nested sampling parameter estimation we now present an example method using bootstrap resampling.

\subsection{Bootstrap estimate of sampling errors}\label{sec:bootstrap}

Given $n$ observations $\mathbf{x}=(x_1,\dots,x_n)$, the bootstrap~\citep{Efron1979} creates new data sets $\mathbf{x}^{\ast}_b$ by drawing $n$ samples from $\mathbf{x}$ with replacement.
This corresponds to approximating the probability distribution of a single data point $x$ as
\begin{equation}
    P(x) \approx \frac{1}{n} \sum^n_{i=1} \delta(x - x_i),
\end{equation}
where $\delta(x)$ is the Dirac delta function~\citep{Ivezic2014}.

As the form of the distribution of sampling errors for a general nested sampling parameter estimation problem is not known, we use the non-parametric bootstrap.
In this case the uncertainty on a quantity $T(\mathbf{x})$ calculated from the data can be estimated by calculating $T(\mathbf{x}_b^\ast)$ for a number of resampled data sets $b = 1,\dots,B$.
For example the bootstrap estimate of the standard error on $T(\mathbf{x})$ is
\begin{equation}
    \std[T(\mathbf{x})]=\sqrt{\frac{1}{B-1}\sum^B_{b=1} {\left(T(\mathbf{x}^{\ast}_b) - \overbar{T(\mathbf{x}_b^\ast)} \,\right)}^2},
    \quad \mathrm{where} \,\,
    \overbar{T(\mathbf{x}_b^\ast)} = \frac{1}{B} \sum^B_{b=1} T(\mathbf{x}^{\ast}_b).
    \label{equ:bootstrap_variance}
\end{equation}

There are many methods for calculating approximate credible intervals on $T(\mathbf{x})$ from bootstrap replications $\{T(\mathbf{x}^\ast_b)\}$ --- see~\citet{Efron1986} for a detailed discussion.
A simple approach from \citet{Johnson2001} is to estimate the boundaries of the $100\alpha\% $ and $100(1-\alpha)\% $ credible regions\footnote{If the distribution of bootstrap replications $T(\bx^\ast_b)$ is skewed then the implied probability distribution of $T$ is skewed in the opposite direction, as can be seen from~\eqref{equ:bootstrap_conf_lower} and~\eqref{equ:bootstrap_conf_upper}. See~\citet[][Section 2]{Loredo2012} for a discussion.} as
\begin{eqnarray}
    {\mathrm{C.I.}}_{100\alpha\%}\left(T(\bx)\right)
    &=&
    2T(\bx) - G^{-1}(1 - \alpha )\label{equ:bootstrap_conf_lower} \\
    {\mathrm{C.I.}}_{100(1- \alpha)\%} \left(T(\bx)\right)
    &=&
    2T (\bx) - G^{-1}(\alpha),\label{equ:bootstrap_conf_upper}
\end{eqnarray}
where $G^{-1}(x)$ is the inverse cumulative distribution of the bootstrap samples $\{T(\mathbf{x}^\ast_b)\}$.
$B=50$ is typically sufficient for an estimate of the standard deviation of a parameter estimate due to sampling errors, but depending on the method used credible intervals on parameter estimates may require 1,000 bootstrap replications or more~\citep{Efron1986}.

When the bootstrap is applied to nested sampling each observation $x_i$ is a thread, and the number of observations is $n$.
Calculating the quantity $T(\mathbf{x})$ involves first combining the set of threads $\mathbf{x}$ into a single run using~\citet{Skilling2006}'s method (described in Section~\ref{sec:divide}), then performing a standard nested sampling calculation including estimating the weight of each point $\wit$ statistically.
Including the same thread multiple times does not cause problems --- repeated dead points $\theta_i = \theta_{i+1}$ are simply assigned the weights $w_i(\bt)$ and $w_{i+1}(\bt)$ respectively.

The following algorithm provides a set of bootstrap replications and an estimate of the standard deviation of sampling errors.

\begin{algorithm}[H]\SetAlgoLined{}
    \caption{Bootstrap sampling error calculation.}                                                                                                                                                      
    \KwResult{Sampling errors and bootstrap replications for the nested sampling calculation $T(\mathrm{dead\,points,weights})$.}
    \KwData{List of dead points and the steps they were sampled at.}
    Divide dead points into a list of threads $\mathbf{x}$ using Algorithm~\ref{alg:splitting}\;
    \While{$b \in B$}
    {create a list of $n$ threads $\mathbf{x}_b^\ast$ by sampling $\mathbf{x}$ with replacement\;
    calculate $T(\mathbf{x}_b^\ast) \equiv T(\mathrm{dead\,points}^\ast_b,\mathrm{weights}^\ast_b)$\;
    }
    calculate $\std[T(\mathbf{x})]=\sqrt{\frac{1}{B-1}\sum^B_{b=1} {\left(T(\mathbf{x}^{\ast}_b) - \overbar{T(\mathbf{x}_b^\ast)} \,\right)}^2}$.
\end{algorithm}
We find that bootstrap resampling gives better results than jackknife resampling, which fails to calculate sampling errors on credible intervals of posterior distributions of parameters such as $\poconf{84}$.
The Bayesian bootstrap was not used as it gives each observation a non-integer weight, which requires modifying nested sampling's use of dead points to statistically estimate prior volume shrinkages.

Resampling techniques such as the bootstrap can generate many simulated runs with the same number of live points $n$ as the original run.
In comparison sampling error estimates from simply splitting a run into many smaller runs and assessing their variation perform poorly, as shown in Section~\ref{sec:split_analysis} of the supplementary material.

\section{Numerical tests}\label{sec:numerical_tests}

Following \citet{Keeton2011} we first test our new method using analytic cases where uncorrelated samples can be easily obtained from the prior within an iso-likelihood contour, allowing us to perform perfect nested sampling.
This ensures our results are not affected by imperfect implementation of the nested sampling algorithm by a specific software.

\begin{figure}
	\centering
    \includegraphics[width=\linewidth]{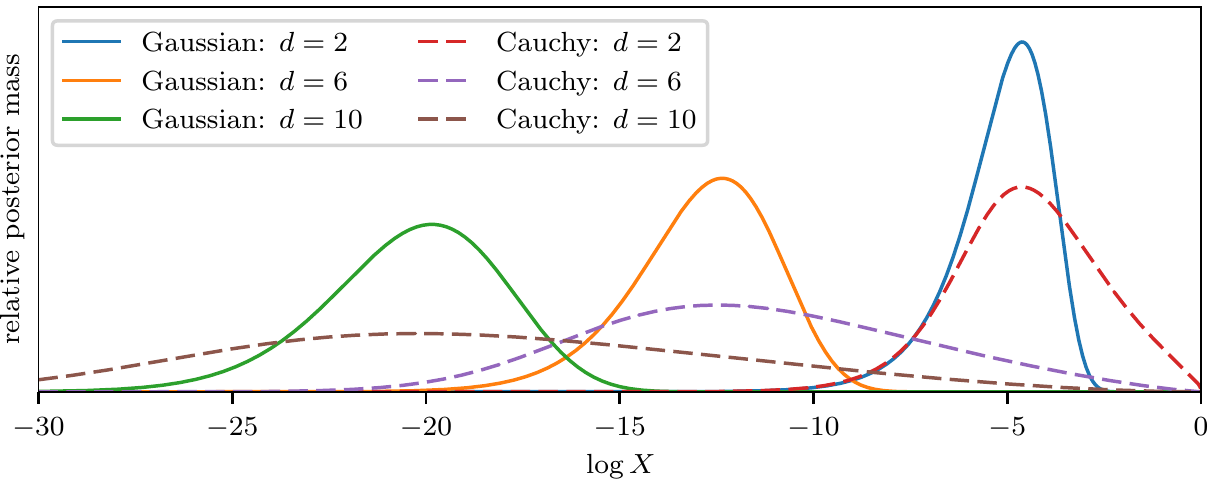}
    \caption{Relative posterior mass as a function of $\logx$ ($\propto \like(X)X$) for Gaussian likelihoods~\eqref{equ:gaussian} and Cauchy likelihoods~\eqref{equ:cauchy} of different dimensions $d$ with Gaussian priors~\eqref{equ:gaussian_prior}. The lines are scaled so that the area under each of them is equal.}\label{fig:an_w_gaussian_cauchy}
\end{figure}

As discussed in~\Cref{sec:theory} perfect nested sampling parameter estimation problems depend on the likelihood $\like(\theta)$ and prior $\pi(\theta)$ only through the distribution of posterior mass $\like(X)$ and the distribution of parameters on iso-likelihood contours $P(f(\theta)|X)$, both of which are functions of both $\like(\theta)$ and $\pi(\theta)$.
We therefore empirically test our method using a wide range of distributions of posterior mass, and examine several functions of parameters $f(\theta)$ in each case.
We construct such tests using Gaussian likelihoods~\eqref{equ:gaussian} and Cauchy likelihoods~\eqref{equ:cauchy} of a variety of dimensions $d$, each with a Gaussian prior~\eqref{equ:gaussian_prior}.
The different distributions of posterior mass for different $d$ are illustrated in~\Cref{fig:an_w_gaussian_cauchy}; the Cauchy distributions have extremely fat tails with significant sample weights throughout the range of $\logx$ values explored.

We use the termination conditions described by \citet[][Section 3.4]{Handley2015b}, stopping when the estimated evidence contained in the live points is less than $10^{-4}$ times the evidence contained in dead points (see Section~\ref{sec:termination} of the supplementary material for a discussion of termination conditions for nested sampling parameter estimation).
Numerical calculations for high-dimensional cases are performed in two dimensions using the technique described in~\Cref{sec:2d_representation}.

As in~\Cref{sec:theory} we denote the first component of the $\theta$ vector as $\theta_{\hat{1}}$, although by symmetry the results will be the same for any component.
$\po$ and $\vart$ are the first and second moments of the posterior distribution of $\theta_{\hat{1}}$, and the one-tailed $Y\%$ upper credible interval $\poconf{Y}$ is the value $\thetaone^\ast$ for which $P(\thetaone<\thetaone^\ast|\like,\pi)=Y/100$.

\subsection{3-dimensional Gaussian example}

\begin{SCfigure}
  \centering
  \includegraphics[width=0.65\linewidth]{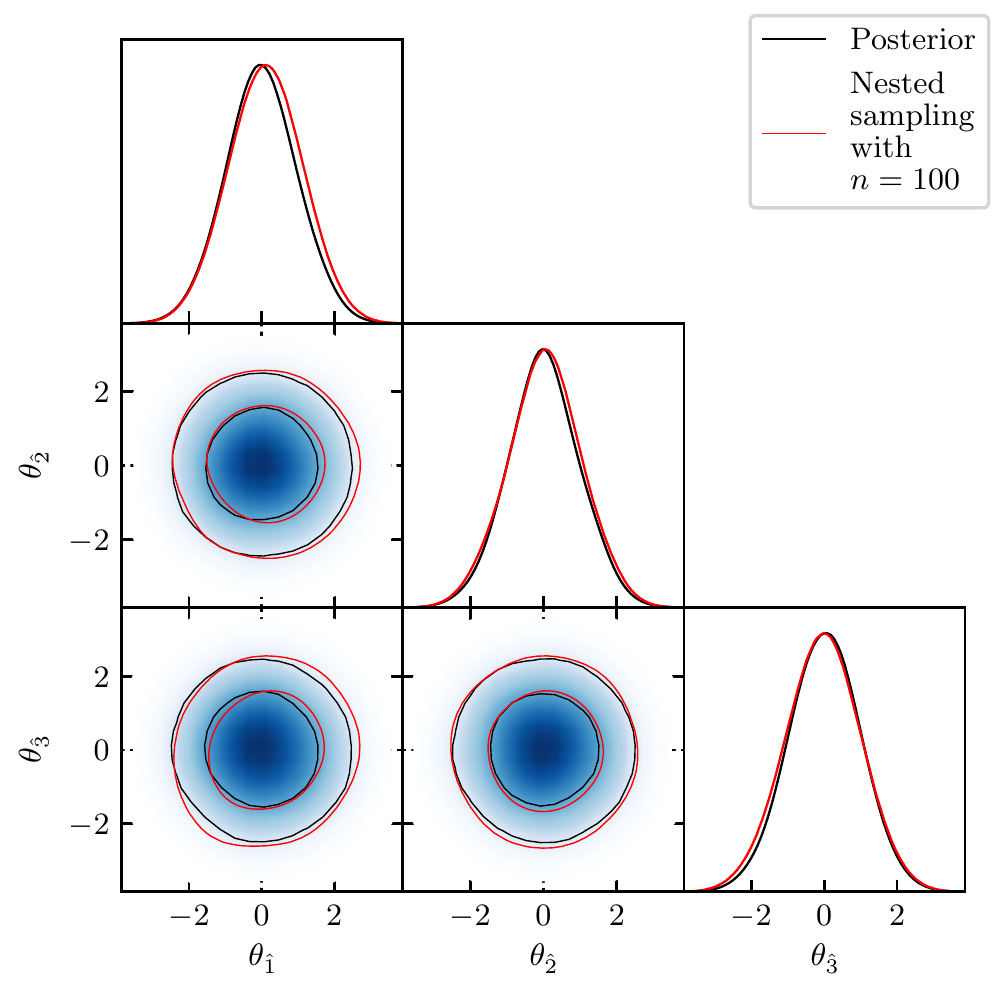} 
  \caption{Sampling errors in a perfect nested sampling calculation for a 3-dimensional Gaussian likelihood~\eqref{equ:gaussian} and a uniform prior.
The shading and black lines show the analytic posterior distribution and the $68\%$ and $95\%$ credible intervals.
The red lines show the calculated posterior credible intervals for a nested sampling run with $n=100$, and differ from the analytic answer due to sampling errors.}%
\label{fig:getdist_gaussian}
\end{SCfigure}

We first test our bootstrap approach to estimating sampling errors on a 3-dimensional Gaussian likelihood~\eqref{equ:gaussian} --- \Cref{fig:getdist_gaussian} illustrates sampling errors on posterior distributions of parameters $\theta$ in this case.
Unlike the simulated weights method, the mean estimates of sampling errors from our method are very close to measurements of sampling errors from repeated calculations --- this shown in the second row of Table~\ref{tab:bootstrap}.
Furthermore the fractional variation of estimates from single runs around the mean estimate is similar to that from the simulated weights method, as shown in the fourth and fifth rows, indicating our method will give a reasonable estimate of sampling errors when only a single nested sampling run is available.

\begin{figure}
	\centering
    \includegraphics[width=0.9\linewidth]{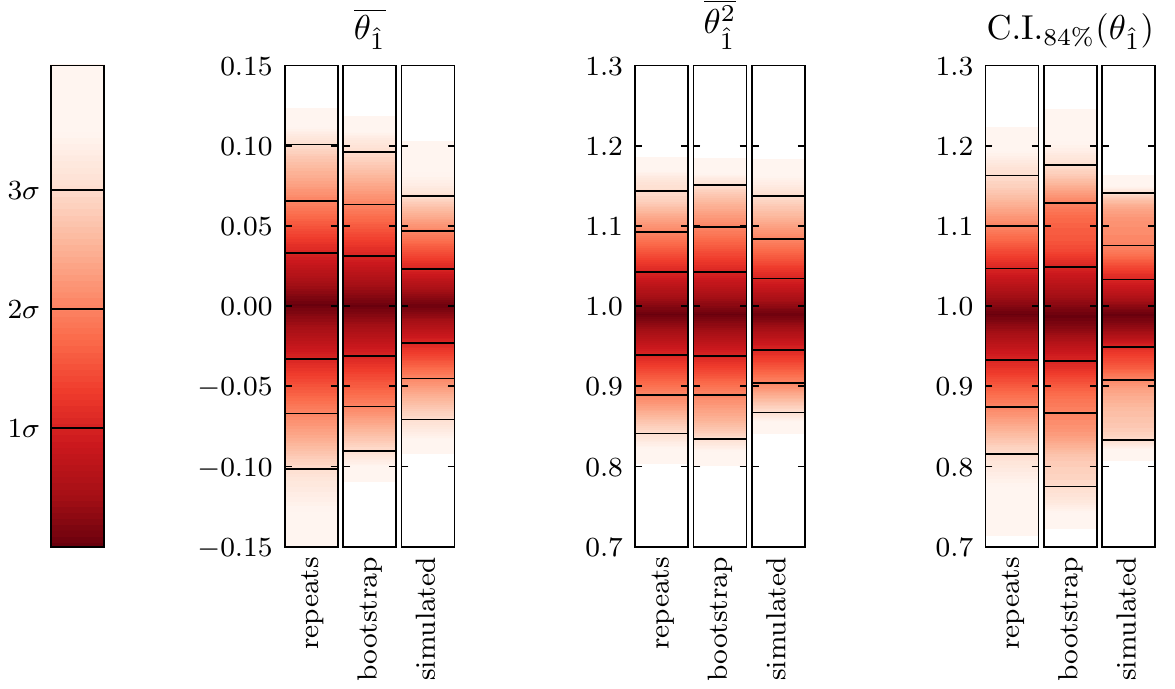}
    \caption{Estimated distributions of sampling errors for parameter estimation with a 3-dimensional Gaussian likelihood~\eqref{equ:gaussian} for perfect nested sampling with $n=200$.
For each estimator the first plot uses values from 5,000 nested sampling runs; the second and third plot are calculated from a single nested sampling run and use 5,000 simulated weights and bootstrap replications. The bootstrap distributions are calculated using~\eqref{equ:bootstrap_conf_lower} and~\eqref{equ:bootstrap_conf_upper}.
The simulated weights and bootstrap values were adjusted by subtracting the difference between their run's expected value for each estimator and its analytical value to line up the distributions.
\Explaincs{}
}\label{fig:bootstrap_color_dists}
\end{figure}

The final two rows of~\Cref{tab:bootstrap} show the empirical coverage rates for bootstrap credible intervals are very close to their nominal values.
Figure~\ref{fig:bootstrap_color_dists} shows estimates of the full sampling error distribution from a single run nested sampling run using the bootstrap and simulated weights methods; the bootstrap results are much closer to the sampling errors observed in repeated calculations, and give accurate estimates of the $1\sigma$ and $2\sigma$ credible intervals.

Section~\ref{sec:additional_tests} of the supplementary material shows similar numerical tests for a 3-dimensional Cauchy likelihood~\eqref{equ:cauchy}.
Even for this challenging, fat-tailed distribution our method performs similarly to the Gaussian case, giving accurate mean error estimates and estimates of credible intervals with measured coverage similar to their nominal coverage.

\begin{table}
\begin{tabular}{l c c c}
    \toprule
    & $\po$ & $\vart$ & $\poconf{84}$ \\
    \midrule
    Repeated runs \std{}                        & $ 0.032(0.2)               $ & $ 0.050(0.4)               $ & $ 0.055(0.4)               $\\
    Bootstrap \std{} / Repeats \std{}           & $ 1.003(7)                 $ & $ 0.998(7)                 $ & $ 1.008(8)                 $\\
    Simulated $w_i$ $\std$ / Repeats $\std$     & $ 0.715(5)                 $ & $ 0.882(6)                 $ & $ 0.785(7)                 $\\
    Bootstrap $\std$ estimate variation         & $ 7.5(1)\%                 $ & $ 8.6(1)\%                 $ & $ 17.7(3)\%                $\\
    Simulated $w_i$ estimate variation          & $ 6.0(1)\%                 $ & $ 7.3(1)\%                 $ & $ 19.7(3)\%                $\\
    Bootstrap $\mathrm{C.I.}_{95\%}$            & $ 0.053(3)                 $ & $ 1.080(5)                 $ & $ 1.077(7)                 $\\
    Bootstrap Mean$\pm1\std$ coverage           & $ 68.4\%                   $ & $ 68.2\%                   $ & $ 68.9   \%                $\\
    Bootstrap $\mathrm{C.I.}_{95\%}$ coverage   & $ 95.0\%                   $ & $ 93.4\%                   $ & $ 93.1   \%                $\\
    \bottomrule
\end{tabular}
    \caption{Sampling errors for a 3-dimensional Gaussian likelihood~\eqref{equ:gaussian}, a Gaussian prior~\eqref{equ:gaussian_prior} and $n=200$.
The first row shows the standard deviation of 10,000 nested sampling calculations.
The second and third rows show the mean of 2,000 error estimates from the bootstrap and simulated weights methods respectively as a ratio to the error observed from repeated calculations; $200$ weight simulations and $200$ bootstrap replications were used for each run.
The fourth and fifth rows show the standard deviations of sampling error estimates for both methods as a percentage of the mean estimate.
The sixth row shows the mean of 100 bootstrap estimates of the one-tailed $95\%$ credible interval on the calculation result given the sampling error, each using 1,000 bootstrap replications.
The final two rows show the empirical coverage of the bootstrap standard error and $95\%$ credible interval from the 10,000 repeated calculations.
\Explainbrackets{}
}\label{tab:bootstrap}
\end{table}

\subsection{Sampling errors in different dimensions}

\begin{figure}
\centering
\subfigure[Parameter estimation sampling errors for Gaussian likelihoods~\eqref{equ:gaussian}.]{\includegraphics[width=\linewidth]{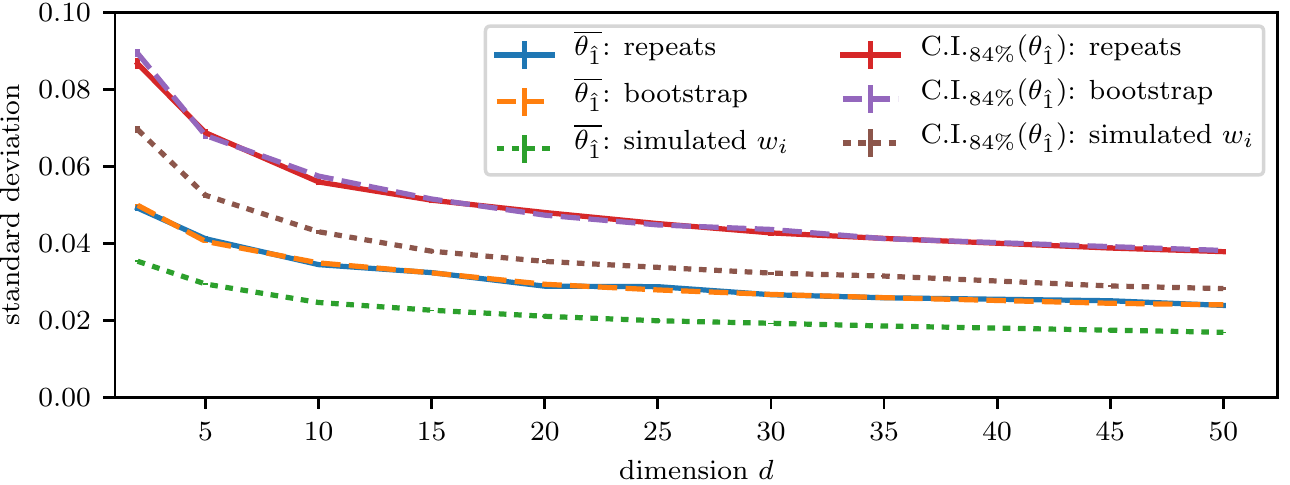}\label{sub:plz_gaussian}}\\
\subfigure[Parameter estimation sampling errors for Cauchy likelihoods~\eqref{equ:cauchy}.]{\includegraphics[width=\linewidth]{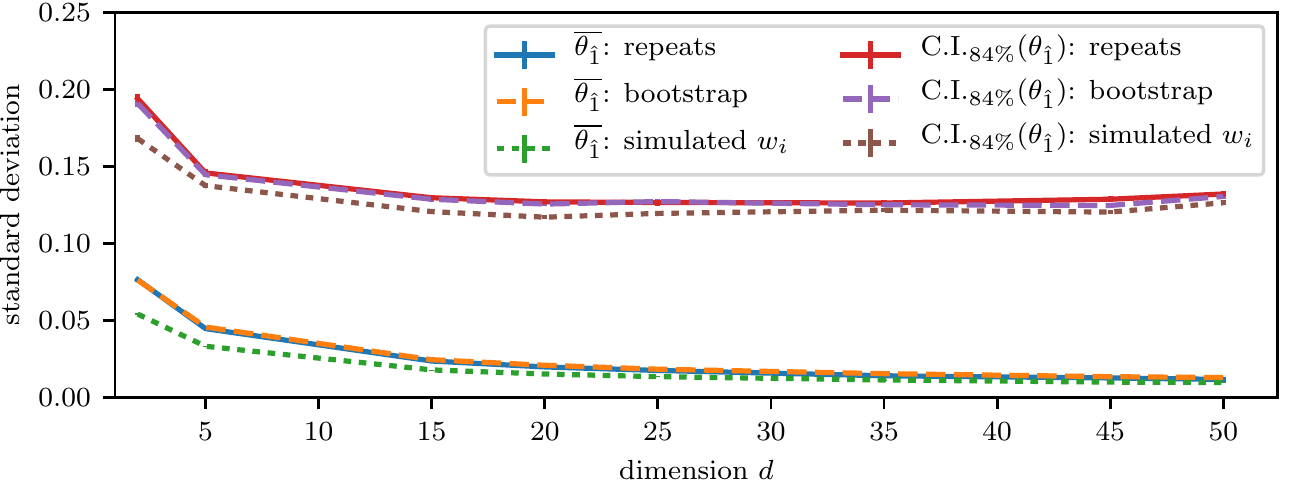}\label{sub:plz_cauchy}}\\
\subfigure[Evidence sampling errors for Gaussian~\eqref{equ:gaussian} and Cauchy~\eqref{equ:cauchy} likelihoods.]{\includegraphics[width=\linewidth]{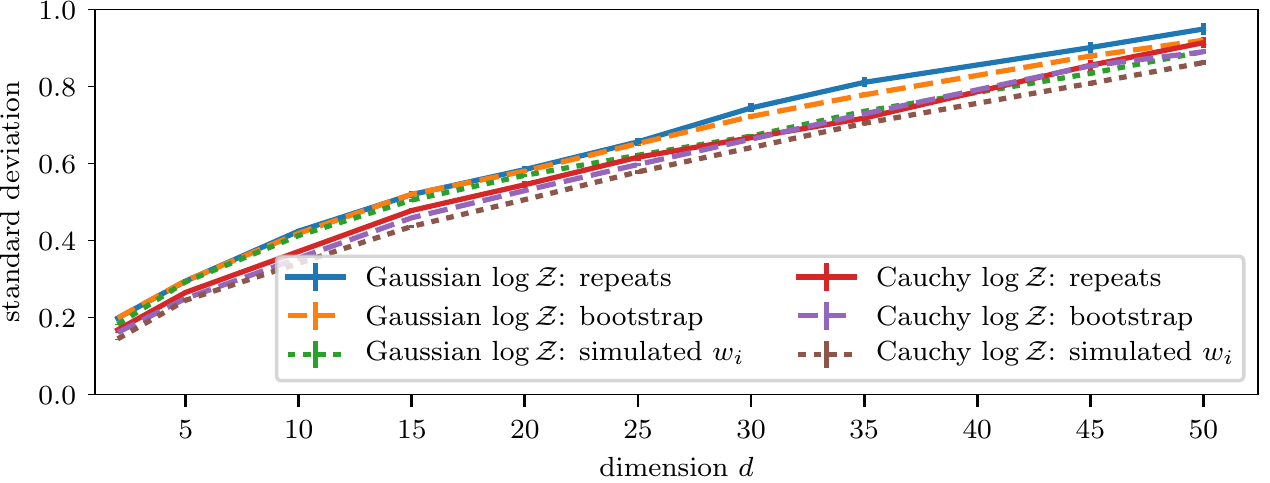}\label{sub:zlr}}\\
\caption{Sampling errors for likelihoods of different dimensions $d$; all use use $d$-dimensional Gaussian priors~\eqref{equ:gaussian_prior} and $n=100$.
Solid lines show the standard deviation of the results of 2,000 calculations.
Dashed and dotted lines show the mean of 500 standard error estimates using the bootstrap and simulated weights methods respectively.
}\label{fig:line_results}
\end{figure}

We now verify the bootstrap method's accuracy for Gaussian~\eqref{equ:gaussian} and Cauchy~\eqref{equ:cauchy} likelihoods of between 2 and 50 dimensions.
\Cref{fig:line_results} shows bootstrap sampling error estimates accurately match the errors measured from repeated calculations, even for the challenging fat-tailed Cauchy distribution.
In contrast the simulated weights method consistently underestimates the sampling errors in parameter estimation, although as expected it is accurate for errors on the evidence $\logZ$.
See Section~\ref{sec:analytical_sqrt_2} of the supplementary material for a detailed discussion of the simulated weights method.

As the dimension $d$ increases, \Cref{fig:line_results} shows parameter estimation errors decreasing and the evidence errors increasing (with a constant number of live points $n$).
This effect is due to the posterior being contained in a smaller fraction of the prior volume in higher dimensions.
In the spherically symmetric cases considered, the range of $\logx$ to be explored increases approximately linearly with the dimension $d$, as can be seen in~\Cref{fig:an_w_gaussian_cauchy}.
With a constant number of live points, the number of samples is therefore also approximately proportional to $d$.

In parameter estimation from posterior samples only points' relative weights matter, so the increased number of samples in higher dimension problems typically increases accuracy as can be seen in~\Cref{sub:plz_gaussian,sub:plz_cauchy}.
However for high dimensional Cauchy likelihoods~\eqref{equ:cauchy} the posterior mass is spread over a wide range of $\logx$ values, so errors in the relative weights of points become large in high dimensions\footnote{When the errors in points' relative weights become dominant the simulated weights method captures the majority of the sampling error, as can be seen for high dimensional Cauchy distributions in~\Cref{sub:plz_cauchy}}.

For $\logZ$ the dominant error is in the absolute value of point weights, which is approximately proportional to the square root of the number of steps required to reach the posterior~\citep{Skilling2006}. $\logZ$ errors are therefore approximately proportional to $\sqrt{d}$ when $n$ is constant, as can be seen in~\Cref{sub:zlr}.

\section{Application to existing nested sampling software}\label{sec:implementation_specific}

Nested sampling software such as \MultiNest{} and \PolyChord{} can be easily modified to output information about the step at which dead points were sampled and give sampling error estimates using bootstrap resampling of threads. We are currently working on incorporating this into future releases of \PolyChord{}.

Sampling error estimates from our approach will be accurate provided the software is performing nested sampling approximately correctly.
However such software can only approximately sample randomly from the prior within iso-likelihood contours --- this may result in additional errors which are specific to a given implementation and which may not be captured by general methods such as resampling threads.
Tuning parameters such as ``\texttt{num\_repeats}'' (the number of slice samples taken between dead points) in \PolyChord{} allow reduced correlation between samples at a higher computational cost.
Testing sampling error estimates from our method against those from repeated calculations can be used to detect implementation-specific errors and to select appropriate values for tuning parameters.

In addition our algorithm for dividing nested sampling runs could be used to detect implementation-specific errors by testing the difference in correlation between threads from the same run and from different runs. In principle this could allow estimates of sampling errors to be corrected for the effects of correlations between threads.

We now demonstrate our method's application to nested sampling results produced with \PolyChord{}.

\subsection{Sampling errors on data fitting with \PolyChord{}}

\begin{figure}[]
    \centering
    \begin{minipage}{0.48\textwidth}
    \includegraphics[width=\linewidth]{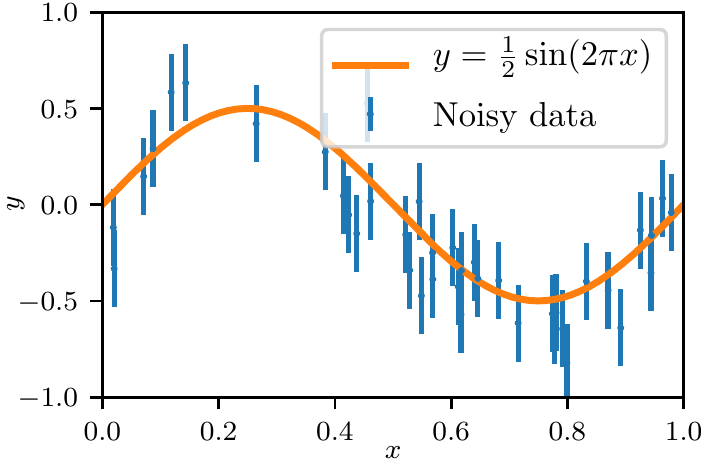}
    \caption{The true $y(x)$ function and the 40 data points used in numerical tests. Data has Gaussian noise of size $\sigma_y = 0.2$.}\label{fig:sin_data}
    \end{minipage}
    \hfill
    \begin{minipage}{0.48\textwidth}
    \includegraphics[width=\linewidth]{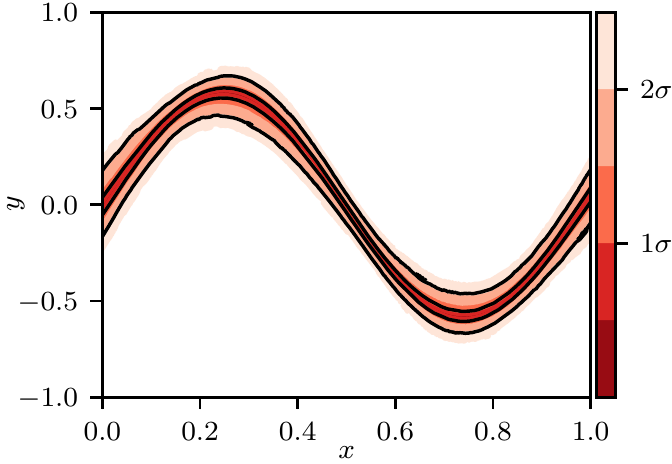}
    \caption{Posterior distribution on $y(x)$ given the data $\mathcal{D}$. The color scale indicates credible intervals.}\label{fig:sin_fgivenx}
    \end{minipage}
\end{figure}

\begin{SCfigure}
  \centering
  \includegraphics[width=0.65\linewidth]{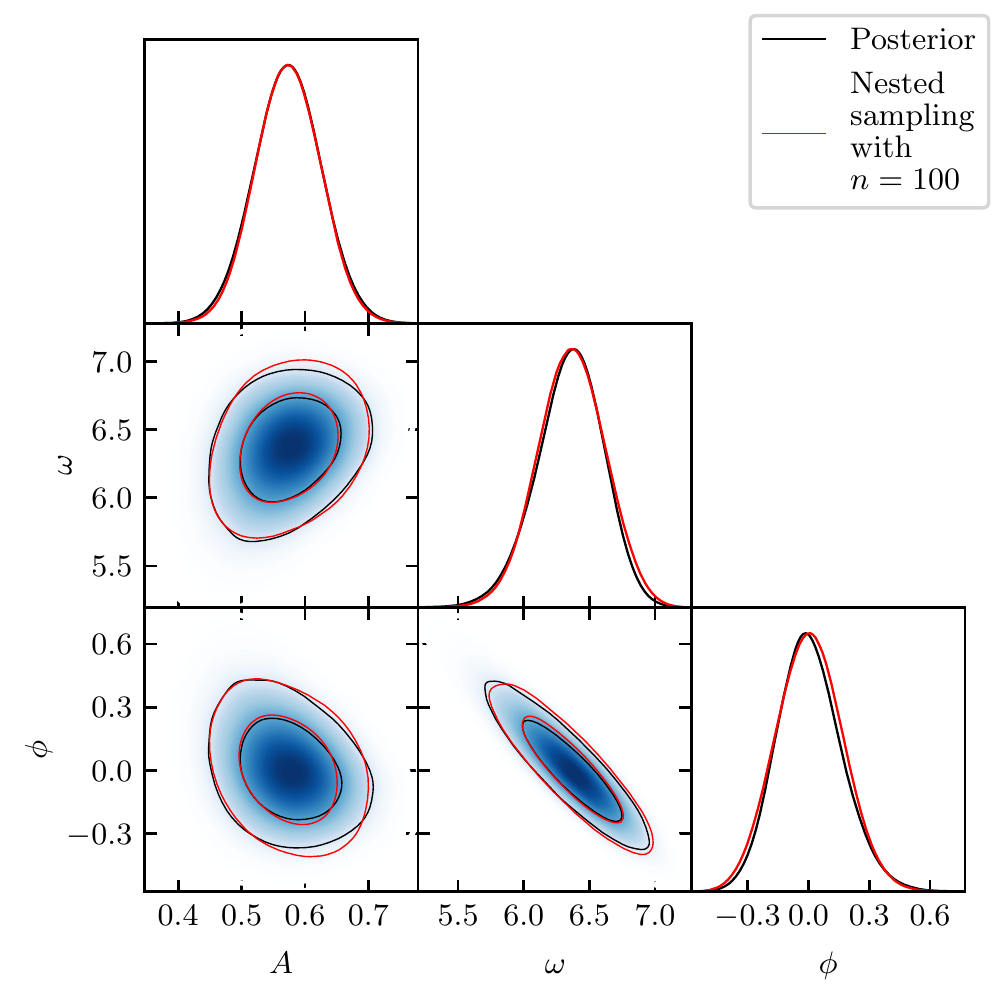} 
  \caption{Posterior distributions and sampling errors from fitting a sinusoid to data~\eqref{equ:fitting_likelihood} using \PolyChord{}.
  The shading and black lines show an accurate calculation of posterior distribution and the $68\%$ and $95\%$ credible intervals from combining 1,000 nested sampling runs with $n=100$.
The red lines show the calculated posterior credible intervals for a single nested sampling run with $n=100$, and differ from the true posterior due to sampling errors.}\label{fig:getdist_sin}
\end{SCfigure}

We fit a set of points $\mathcal{D}=\{x_i, y_i\}$ with normally distributed errors $\sigma_y$ on the $y$ values using a sinusoid
\begin{equation}
y(x) = A \sin(\omega x + \phi).
\end{equation}
The likelihood is then
\begin{equation}
    \like(\theta) = \prod_i \frac{1}{\sqrt{2 \pi {\sigma_y}^2}} \e^{-{\left(y_i - y(x)\right)}^2 / 2 \sigma_y^2},
    \label{equ:fitting_likelihood}
\end{equation}
where $\theta=(A,\omega , \phi )$ and we use a uniform prior for $A \in (0,1)$, $\omega \in (0, 10)$ and $\phi \in (-\pi/2, \pi/2)$.
Numerical tests use 40 data points sampled from $y(x) = \frac{1}{2} \sin( 2 \pi x )$ with Gaussian noise of size $\sigma_y =0.2$ added to the $y$ values; $y(x)$ and the data points are shown in~\Cref{fig:sin_data} and the posterior distribution of $y(x)$ given the data is shown in~\Cref{fig:sin_fgivenx}.
Posterior distributions on $A$, $\omega$ and $\phi$ can be calculated with nested sampling --- these are illustrated in~\Cref{fig:getdist_sin} along with example sampling errors.

Table~\ref{tab:polychord} shows sampling errors from \PolyChord{} with $\texttt{num\_repeats}=15$ --- the default value for a 3-dimensional problem.
As in perfect nested sampling, our bootstrap estimates of the standard error agree with the variation in results observed, and the observed coverage of credible intervals is close to their nominal coverage.
This implies that $\texttt{num\_repeats}=15$ is sufficient for \PolyChord{} to perform parameter estimation accurately in this case.

\begin{table}
	\centering
	\begin{tabular}{l c c c}
    \toprule
    & $A$ & $\omega$ & \conf{\omega}{84} \\
    \midrule
	Repeated runs \std{}                        & $ 0.247(6) \cdot 10^{-2}   $ & $ 0.012(0.3)               $ &  $ 0.019(0.4)               $ \\
	Bootstrap \std{} / Repeats \std{}           & $ 0.98(2)                  $ & $ 0.98(2)                  $ &  $ 0.97(2)                  $ \\
	Simulated $w_i$ $\std$ / Repeats $\std$     & $ 0.71(2)                  $ & $ 0.71(2)                  $ &  $ 0.73(2)                  $ \\
	Bootstrap $\std$ estimate variation         & $ 8.8(2)\%                 $ & $ 9.3(2)\%                 $ &  $ 19.1(4)\%                $ \\
	Simulated $w_i$ estimate variation          & $ 6.4(1)\%                 $ & $ 6.6(1)\%                 $ &  $ 21.5(5)\%                $ \\
	Bootstrap $\mathrm{C.I.}_{95\%}$            & $ 0.577(0.2)               $ & $ 6.370(0.8)               $ &  $ 6.632(1)                 $ \\
	Bootstrap Mean$\pm1\std$ coverage           & $ 69.4\%                   $ & $ 69.3\%                   $ &  $ 66.9\%                   $ \\
	Bootstrap $\mathrm{C.I.}_{95\%}$ coverage   & $ 93.4\%                   $ & $ 95.0\%                   $ &  $ 95.1\%                   $ \\
    \bottomrule
\end{tabular}
\caption{Sampling errors for the sinusoid fitting likelihood~\eqref{equ:fitting_likelihood} using \PolyChord{} with $n=100$.
The first row shows the standard deviation of 1,000 nested sampling calculations.
The second and third rows show the mean of 1,000 error estimates from the bootstrap and simulated weights methods respectively as a ratio to the error observed from repeated calculations; $200$ weight simulations and $200$ bootstrap replications were used for each run.
The fourth and fifth rows show the standard deviations of sampling error estimates for both methods as a percentage of the mean estimate.
The sixth row shows the mean of 100 bootstrap estimates of the one-tailed $95\%$ credible interval on the calculation result, each using 1,000 bootstrap replications.
The final two rows show the empirical coverage of the bootstrap standard error and $95\%$ credible interval from the 1,000 repeated calculations.
\Explainbrackets{}
}\label{tab:polychord}
\end{table}

\section{Conclusion}

Sampling errors in nested sampling parameter estimation arise principally from two sources: uncertain sample weights $\wit$, and approximating the average of a function of parameters on each iso-likelihood contour $\ftilde(X_i)$ with a single sample $f(\theta_i)$.
The latter error is not present in evidence calculation and has been previously ignored.
The added stochasticity from sampling each iso-likelihood contour makes nested sampling parameter estimation a 2-dimensional problem, with a dependence on both the distribution of posterior mass $\like(X)$ and the distribution of parameter values $P(f(\theta)|X)$ on each iso-likelihood contour.
We proposed a new diagram for representing both aspects of the calculation, and presented it in~\Cref{sub:p1_gaussian,sub:p1_gaussian_3d,sub:p1_cauchy,sub:p1var_gaussian,sub:r_gaussian}.

Estimating sampling errors is vital for interpreting the results of a nested sampling calculation, as well as for allocating computational resources --- for example by choosing an appropriate number of live points.
However the current approach (the simulated weights method) underestimates sampling errors as it does not account for approximating $\ftilde(X_i)$ with a single sample $f(\theta_i)$.
We proposed a new method for estimating sampling errors using our new algorithm (Algorithm~\ref{alg:splitting}) for dividing a nested sampling run into single live point runs (``threads''), which can then be resampled with techniques such as the bootstrap.
This works as the $\log X_i$ values of the dead points $i$ from some nested sampling run with $n$ live points form a Poisson process with rate $n$, meaning the $\log X_j$ values of the dead points $j$ of a single thread are a Poisson process of rate 1.

Our method shows accurate and robust estimation of sampling errors in parameter estimation in empirical tests, and compares favourably to the other methods discussed.
The new method can be easily incorporated into existing nested sampling software, and will be reliable provided the implementation is performing the nested sampling algorithm accurately.
We are currently working on including nested sampling run division and sampling error estimates from our method in future versions of \PolyChord{}.

\bibliographystyle{ba}
\bibliography{library}

\begin{acknowledgement}

The authors are grateful to the referees and editors for their helpful comments and suggestions, which have greatly improved the paper.
In particular we thank one of the reviewers for providing the justification of resampling nested sampling threads in terms of Poisson processes.

\end{acknowledgement}
\clearpage

\begin{appendices}

\section*{Supplementary material} 

\section{Relative contributions of different sources of parameter estimation sampling errors}\label{sec:relative_errors}

\begin{figure}[!]
\centering
    \includegraphics[width=0.9\linewidth]{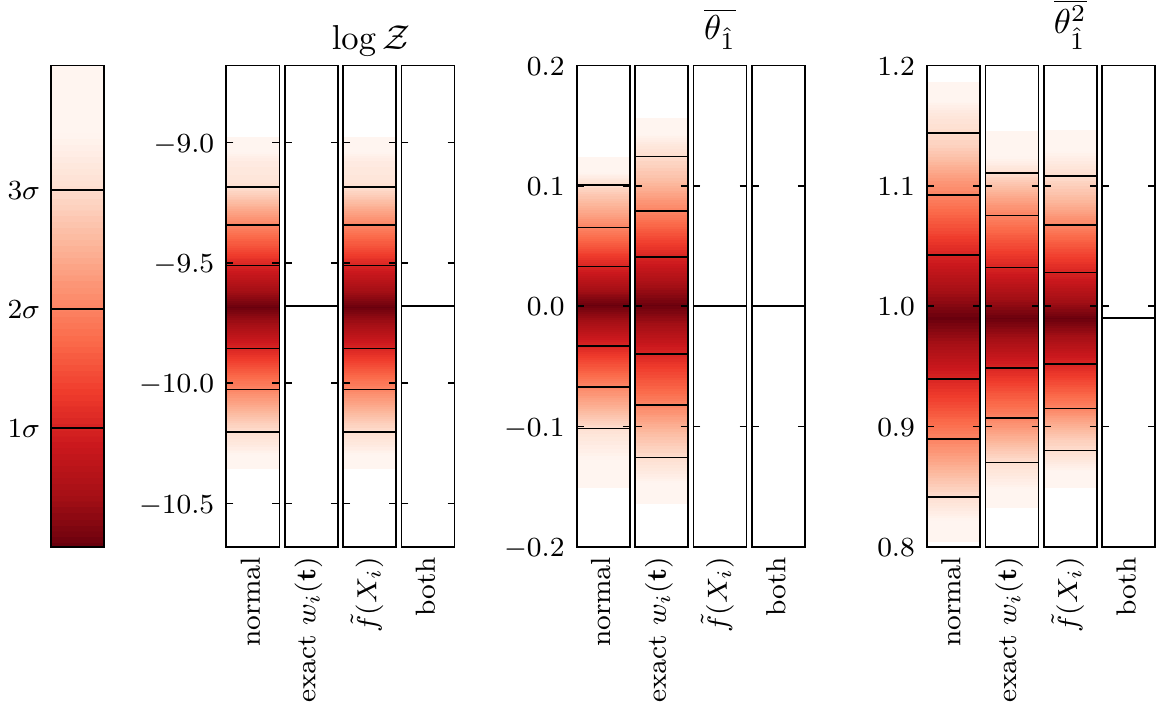}
    \caption{Sources of sampling error in perfect nested sampling with a 3-dimensional Gaussian likelihood~\eqref{equ:gaussian}, a Gaussian prior~\eqref{equ:gaussian_prior} and $n=200$.
Each plot shows the distribution of the results of 5,000 nested sampling calculations.
For each estimator the first bar is from standard nested sampling, the second bar uses analytically calculated prior volumes for its sample weights $\wit$ and the third bar uses $\ftilde(X_i)$ instead of $f(\theta)$ to calculate estimates.
The fourth bar uses both analytical $\wit$ and $\ftilde(X_i)$ values --- the error in this case is very small and calculation results are all close to the analytic answer.
\Explaincs{}                                                                                                                      
}%
\label{fig:error_types}
\captionsetup[table]{labelsep=space} 
\begin{tabular}{l c c c}
    \toprule
    & \stdct{\logZ} & \stdct{\po} & \stdct{\vart} \\
    \midrule
    Normal runs 					 & $ 0.169(2)               $ & $ 0.033(3)                $ & $ 0.051(0.5)                $\\
    Exact $\wit$ 					 & $ 0.394(4) \cdot 10^{-5} $ & $ 0.040(4)                $ & $ 0.040(0.4)                $\\
    Sampling $\ftilde(X_i)$ 		 & $ 0.169(2)               $ & $ 0.000(0)                $ & $ 0.038(0.4)                $\\
    Exact $\wit$ and $\ftilde(X_i)$  & $ 0.394(4) \cdot 10^{-5} $ & $ 0.000(0)                $ & $ 0.330(3) \cdot 10^{-5} $\\
    \bottomrule
\end{tabular}
    \captionof{table}{The standard deviations of the sampling error distributions in Figure~\ref{fig:error_types}; \explainbrackets{}
For $f(\theta)=\thetaone$, $\ftilde(X)=0$ for all $X(\theta)$ and so when $\ftilde(X_i)$ values are used every calculation gives $\po=0$ without any sampling error.}%
\label{tab:error_types}
\captionsetup[table]{labelsep=colon} 
\end{figure}

The relative contributions of sampling errors from unknown prior weights of points $\wit$ and from taking a single sample $\theta_i$ on each iso-likelihood contour (discussed in~\Cref{sec:theory}) can be calculated by using exact values for weights $\wit$ and replacing $f(\theta_i)$ with $\ftilde(X_i)$.
For a Gaussian likelihood~\eqref{equ:gaussian} and Gaussian prior~\eqref{equ:gaussian_prior} both $\wit$ and $\ftilde(X_i)$ can be calculated analytically for each $\theta_i$ --- sampling errors from calculations using this additional information are shown in Figure~\ref{fig:error_types} and Table~\ref{tab:error_types}.

When calculating $\logZ$, as expected, using exact weights $\wit$ reduces uncertainty to the small trapezium rule error and using $\ftilde(X_i)$ has no effect.
However for parameter estimation significant error remains when using exact $\wit$ values%
\footnote{For $f(\theta)=\theta_{\hat{1}}$ the error increases when exact $\wit$ values are used.
This is because the true weights are more variable than the expected ones and this reduces the information content (entropy) of the set of samples.}.
The relative contribution to sampling errors from estimating weights statistically is greatest when $\ftilde(X)$ has a strong dependence on $X$ over the interval in $X$ containing the bulk of the posterior mass.
In contrast when $f(\theta)=\theta_{\hat{1}}$ then $\ftilde(X)=0$ for all $\theta$, and the analysis using $\ftilde(X_i)$ always gives the analytically correct answer of zero.
In all cases, when both exact $\wit$ and samples from $\ftilde(X_i)$ are used the sampling error is reduced to close to zero.

\section{Analysis of the simulated weights method}\label{sec:analytical_sqrt_2}

The simulation method underestimates sampling errors in nested sampling parameter estimation, as shown by the numerical tests in~\Cref{tab:bootstrap,tab:polychord} and~\Cref{fig:line_results}.
This is because it assumes that for each dead point $f(\theta_i) \approx \ftilde(X_i)$, neglecting the sampling errors from taking a single sample on each iso-likelihood contour which are described in Section~\ref{sec:theory}.
However some of this error is captured because repeatedly simulating points' weights behaves like a resampling scheme, with similarities to the Bayesian bootstrap~\citep{Rubin1981}.
Resampling estimates the uncertainty on inferences from a set of samples by calculating its variation when data points are re-weighted, but the simulated weights method does so in a way that systematically underestimates sampling errors.
This behavior has not been documented in the literature.

For example, consider the case $f(\theta)=\theta_{\hat{1}}$ with a Gaussian likelihood~\eqref{equ:gaussian} and Gaussian prior~\eqref{equ:gaussian_prior} --- here $\ftilde(X)=0$ for all $X(\theta)$.
If $\ftilde(X_i)$ is used instead of $f(\theta_i)$ there is no sampling error on estimates of $\overbar{\thetaone}$ regardless of any uncertainty in the weights of each point $p_i$, as can be seen in~\Cref{fig:error_types} and Table~\ref{tab:error_types}.
However the simulated weights method gives a non-zero estimate which on average differs from the sampling errors measured by repeated calculations by a factor of very close to $2^{-\frac{1}{2}} = 0.707$, as shown in the third row of Table~\ref{tab:bootstrap}.

Further numerical tests show that in special cases when $\ftilde(X)$ is the same at all $X$ the ratio of sampling errors from the simulated weights method to the error observed in repeated calculations has a value close to $2^{-\frac{1}{2}}$.
We give an analytical explanation for this result below.
However we note that for practical problems $\ftilde(X)$ is a priori unknown and likely varies in $X$, meaning the true sampling error cannot be predicted by adjusting estimates from the simulated weights method.

\subsection{Sampling error estimates for special cases when $\ftilde(X)$ is constant for all $X$}

\subsubsection{Variance of sampling error distribution}

Nested sampling calculates the expected value of a function of parameters as $\sum_i p_i f(\theta_i)$.
Here the sampling error is the difference between the exact value of $\E[f(\theta)]$ from the posterior, and is distributed as
\begin{equation}
    \mathrm{sampling \: error} \: \sim \: P\left(\, \sum_{s\in\mathcal{S}} p_s f(\theta_s) - \E[f(\theta)]\right).
\end{equation}
The variance of this distribution provides a measure of the size of the sampling error.
As the nested sampling estimator is unbiased, the variance of the sampling error distribution is equal to the variance of the results of repeated calculations:
\begin{equation}
    \Var\left[\sum_{i} p_i(\bt) f(\theta_i)\right] = \sum_{i,j} \Cov \left[p_i(\bt) f(\theta_i), p_j(\bt) f(\theta_j) \right].
\end{equation}
Expanding and dropping the explicit dependence of $p_i$ and $f_i$ on $\bt$ and $\theta_i$ for brevity gives
\begin{align}
    \begin{split}\label{equ:covar_expansion}
    \Cov\left[p_i f_i, p_j f_j \right]
    =
    \E[p_i]\E[p_j] \Cov[f_i, f_j] +
    \E[p_i]\E[f_j] \Cov[f_i, p_j] + \\
    \E[f_i]\E[p_j] \Cov[p_i, f_j] +
    \E[f_i]\E[f_j] \Cov[p_i, p_j] + \\
    \E[(\Delta p_i)(\Delta p_j)(\Delta f_i) (\Delta f_j)] +
    \E[p_i]\E[(\Delta f_i)(\Delta p_j) (\Delta f_j)] + \\
    \E[f_i]\E[(\Delta p_i)(\Delta p_j) (\Delta f_j)] +
    \E[p_j]\E[(\Delta p_i)(\Delta f_i) (\Delta f_j)] + \\
    \E[f_j]\E[(\Delta p_i)(\Delta f_i) (\Delta p_j)] -
    \Cov[p_i,f_i] \Cov[p_j, f_j],
\end{split}
\end{align}
where $\Delta y \equiv y - \E[y]$.

Each $f_i$ is an independent random variable from the distribution $P(f(\theta)|X_i)$, so the expectation of products of $\Delta p_i \Delta f_j$ are zero for all $i,j$.
Furthermore expectation of products $\Delta f_i \Delta f_j$ and the covariance $\Cov[f_i, f_j]$ are zero for $i \ne j$.

The weights $p_i$ have a dependence on $X$, but in the case $\ftilde(X_i)$ is the same for all $X$ the covariance terms $\Cov[f_i, p_j]$ are also zero for all $i,j$.%
~\eqref{equ:covar_expansion} therefore simplifies to
\begin{equation}
    \sum_i \sum_j \Cov\left[p_i f_i, p_j f_j \right]
    =
    \sum_i \Var\left[p_i f_i\right] +
    \sum_{i\ne j, j} \left[
        \E[f_i]\E[f_j] \Cov[p_i, p_j] \right].
\end{equation}
Expanding the variance term on the right hand side when $\ftilde(X)$ is constant and $f_i$ and $p_i$ are therefore independent gives
\begin{align}\label{equ:special_cov}
    \sum_{i,j} \Cov\left[p_i f_i, p_j f_j \right]
    =
    \sum_i \left[ \E[{p_i}^2] \Var[f_i] \right] +
    \sum_{i, j}
    \left[ \E[f_i]\E[f_j] \Cov[p_i, p_j] \right].
\end{align}

\subsubsection{Simulated weights method variance estimate}

The simulated weights method corresponds to fixing the $f_i$ values while retaining the stochastic dependence of $p_i$ on $\bt$.
This means taking ${\E[f_i]}_\mathrm{sim}=f_i$, ${\Var[f_i]}_\mathrm{sim} = 0$, which combined with~\eqref{equ:special_cov} gives
\begin{equation}
    \mathrm{Var}_{\mathrm{simulated}}
    =
    \sum_{i, j} f_i f_j \Cov[p_i, p_j].
\end{equation}
Taking the expected values for $f_i$ and $f_j$ this becomes
\begin{equation}
    \E[\mathrm{Var}_{\mathrm{simulated}}]
    =
    \sum_{i, j}{\E[f_i]}^2  \Cov[p_i, p_j] + \sum_i \left( \E[{f_i^2}] - {\E[f_i]}^2 \right) \Var[p_i].
\end{equation}
Using that by definition $\sum_i p_i = 1$ so $\sum_{i,j} \Cov[p_i, p_j] = \Var[ \sum_i p_i ] = 0$,
\begin{equation}\label{equ:sim_var}
    \E[\mathrm{Var}_{\mathrm{simulated}}]
    =
    \sum_i \Var[f_i] \Var[p_i].
\end{equation}

In contrast the repeated runs method retains the sampling error of $f_i$ on $\theta_i$ and uses the expected values of the weight $\E[p_i]$.
Hence for a large number of trials ${\E[f_i]}_\mathrm{rep}=\E[f_i]=\ftilde(X_i)$, ${\Var[f_i]}_\mathrm{rep} = \Var[f_i]$ for all $i$.
Subbing into~\eqref{equ:special_cov} gives
\begin{equation}
    \E[\mathrm{Var}_{\mathrm{repeats}}]
    =
    \sum_i \Var[f_i] \E\left[{p_i}^2\right] + \sum_{i, j} {\E[f_i]}^2 \Cov\!\left[ \E[p_i], E[p_j] \right].
\end{equation}
Using that $\sum_{i,j} \Cov[E[p_i], E[p_j]] = \Var[ \sum_i E[p_i] ] = 0$,
\begin{equation}\label{equ:rep_var}
    \E[\mathrm{Var}_{\mathrm{repeats}}]
    =
    \sum_i \Var[f_i] \E\left[{p_i}^2\right].
\end{equation}

\subsubsection{Ratio of simulated weights and repeated runs variance estimates}

Combining equations~\eqref{equ:sim_var} and~\eqref{equ:rep_var} gives the ratio of the simulated weights method and repeated runs variances as
\begin{equation}
    \frac
    {\E[\mathrm{Var}_{\mathrm{simulated}}]}
    {\E[\mathrm{Var}_{\mathrm{repeats}}]}
    =
    \frac
    {\sum_i \Var[f_i] \Var[p_i]}
    {\sum_i \Var[f_i] \E\left[{p_i}^2\right]}
    =
    \frac
    {\sum_i \Var[f_i] \left(\E\left[{p_i}^2\right] - {\E\left[p_i\right]}^2\right)}
    {\sum_i \Var[f_i] E\left[{p_i}^2\right]}.
\end{equation}
If $\Var[P(f(\theta)|X)]$ is the same for all $X$ this simplifies to
\begin{equation}
    \frac
    {\E[\mathrm{Var}_{\mathrm{simulated}}]}
    {\E[\mathrm{Var}_{\mathrm{repeats}}]}
    =
    \frac
    {\sum_i \E\left[{p_i}^2\right] - {\E\left[p_i\right]}^2}
    {\sum_i \E\left[{p_i}^2\right]}.
\end{equation}
By definition the normalised weights $p_i \equiv \frac{w_i(\bt)}{\Z(\bt)}$ so
\begin{align}
    \expect{p_i} &= \expect{w_i} \expect{\Z^{-1}} + \Cov \left[w_i, \Z^{-1} \right], \\
    \expect{p_i^2} &= \expect{w_i^2} \expect{\Z^{-2}} + \Cov \left[w_i^2, \Z^{-2} \right].
\end{align}
Numerical results suggest that for a range of problems $p_i$ and $\Z$ are approximately independent, in which case
\begin{equation}
    \frac
    {\E[\mathrm{Var}_{\mathrm{simulated}}]}
    {\E[\mathrm{Var}_{\mathrm{repeats}}]}
    \approx
    \frac
    {\sum_i \Var[f_i] \left[ (\E\left[{w_i}^2\right] - {\E\left[w_i\right]}^2) + \frac{\Var[\Z^{-1}]}{\expect{\Z^{-2}}}\expect{w_i^2}\right]}
    {\sum_i \Var[f_i] \E\left[{w_i}^2\right]}.
\end{equation}
Typical problems with a large $n$ often also have $\Var\left[\Z^{-1}\right] \ll \expect{\Z^{-2}}$, in which case
\begin{equation}\label{equ:approx_sim_rep_ratio}
    \frac
    {\E[\mathrm{Var}_{\mathrm{simulated}}]}
    {\E[\mathrm{Var}_{\mathrm{repeats}}]}
    \approx
    \frac
    {\sum_i \Var[f_i] \left[ \E\left[{w_i}^2\right] - {\E\left[w_i\right]}^2 \right]}
    {\sum_i \Var[f_i] \E\left[{w_i}^2\right]}.
\end{equation}

\citet{Keeton2011} gives expressions for the weights as\footnote{These formulae omit the trapezium rule and for brevity take $w_i(\bt) = \like_i (X_{i-1} - X_i)$ --- this approximation has little effect on the results.}
\begin{align}
    \expect{w_i} =& \E[\like_i] \frac{1}{n} {\left( \frac{n}{n + 1} \right)}^i, \\
    \expect{w_i^2} =& {\E[\like_i]}^2 \frac{2}{n(n + 1)} {\left( \frac{n}{n + 2} \right)}^i.
\end{align}

For a general likelihood the summation in~\eqref{equ:approx_sim_rep_ratio} cannot be found exactly.
However one can estimate the ratio for each live point
\begin{align}
    \frac
    {\E\left[{w_i}^2\right] - {\E\left[w_i\right]}^2}
    {\E\left[{w_i}^2\right]}
    &=
    \frac
    {\frac{2}{n(n + 1)} {\left( \frac{n}{n + 2} \right)}^i - \frac{1}{n^2} {\left( \frac{n}{n + 1} \right)}^{2i}}
    {\frac{2}{n(n + 1)} {\left( \frac{n}{n + 2} \right)}^i} \\
    &=
    \frac{2 - \frac{n + 1}{n} {\left(1 - \frac{1}{{(n+1)}^2} \right)}^i}{2} \\
    &\approx \frac{1}{2} \quad \mathrm{when} \,\, n \gg 1 \,\, \mathrm{and} \,\, i \ll n^2.
\end{align}
This supports the observation that the ratio of simulated weights method estimates of the standard deviation of stochastic errors to measurements from repeated runs is close to $2^{-1/2}$ for special cases such as calculating the mean of a parameter for spherically symmetric likelihoods with spherically symmetric co-centred priors.

\subsubsection{Ratio in the special case where $\like(X)$ and $P(f(\theta)|X)$ are constant for all $X$}

If the likelihood $\like$ is constant%
\footnote{We assume $\like(X)$ has an infinitesimal slope to give direction to nested sampling's inward iteration.}
throughout the parameter space and $\Var[P(f(\theta)|X)]$ is the same for all $X$ then the likelihood terms in the numerator and denominator of~\eqref{equ:approx_sim_rep_ratio} cancel and the summation can be found exactly.
Furthermore estimates of $\Z$ are very precise in this case as there is no stochastic variation in $\{\like_i\}$, justifying the approximation~\eqref{equ:approx_sim_rep_ratio}.
In this case the ratio is
\begin{align}
    \frac
    {\E[\mathrm{Var}_{\mathrm{simulated}}]}
    {\E[\mathrm{Var}_{\mathrm{repeats}}]}
    &\approx
    \frac
    {\sum_i \E\left[{w_i}^2\right] - {\E\left[w_i\right]}^2}
    {\sum_i \E\left[{w_i}^2\right]} \\
    &=
    \frac
    {\sum_i \left[ \frac{1}{n} {\left( \frac{n}{n + 1} \right)}^i +
    1 - \frac{2}{n(n + 1)} {\left( \frac{n}{n + 2} \right)}^i \right]}
    {\sum_i \frac{2}{n(n + 1)} {\left( \frac{n}{n + 2} \right)}^i} \\
    &=
    \frac{1}{2 + \frac{1}{n}},
\end{align}
where the final step sums the geometric series and neglects terms from the truncation of the sum due to termination of the nested sampling run.

\section{Split runs method}\label{sec:split_analysis}

Instead of spending all available computational resources on a single nested sampling run with $n$ live points, one might consider performing $N$ smaller runs with $n/N$ live points and estimating the sampling error from the variation of the smaller runs --- for example as $1/\sqrt{N}$ times their sample standard deviations.
However this provides a limited number of sub-runs, and does not give accurate credible interval estimates.
Furthermore while sampling errors in nested sampling are typically proportional to $1/\sqrt{n}$, this breaks down when the number of samples is small due to trapezium rule errors in sample weights which are $\mathcal{O}(1/n^2)$.
As a result multiple runs are best analysed by combining them into a single run~\citep{Skilling2006}.

Sampling error estimates from taking the standard deviation of the results of $N=20$ sub-runs and multiplying by $1/\sqrt{N}$ are shown in Table~\ref{tab:split}.
The split runs method is inaccurate for the approximately log-normally distributed sampling errors in $\Z$ as well as for credible intervals on distribution tails such as $\poconf{84}$, as can be seen in the third row of Table~\ref{tab:split}.

\begin{table}
\centering
\begin{tabular}{l c c c}
    \toprule
    & $\Z$ & $\po$ & \poconf{84} \\
    \midrule
    Repeats $\std$                              & $ 0.111(1) \cdot 10^{-4}   $ & $ 0.032(0.2)               $ & $ 0.055(0.4)               $ \\
    Split into 20 $\std$ / Repeats $\std$       & $ 1.332(15)                $ & $ 1.012(8)                 $ & $ 0.972(8)                 $ \\
    Bootstrap $\std$ / Repeats $\std$           & $ 1.009(8)                 $ & $ 1.003(7)                 $ & $ 1.008(8)                 $ \\
    Split into 20 $\std$ estimate variation     & $ 37.9(6)\%                $ & $ 16.4(3)\%                $ & $ 16.1(3)\%                $ \\
    Bootstrap $\std$ estimate variation         & $ 17.6(3)\%                $ & $ 7.5(1)\%                 $ & $ 17.7(3)\%                $ \\
    \bottomrule
\end{tabular}
\caption{Test of the split analysis method using perfect nested sampling with a 3-dimensional unit Gaussian likelihood~\eqref{equ:gaussian}, a Gaussian prior~\eqref{equ:gaussian_prior} and $n=200$.
    The first row shows the standard deviation of results from 10,000 nested sampling calculations.
    The second row shows the mean estimate of sampling error standard deviation from 2,000 individual runs using the split method, breaking each run into 20 smaller runs with $n = 10$.
    The third row shows the mean of 2,000 bootstrap estimates of the sampling errors for comparison.
    The fourth and fifth row shows the standard deviation of error estimates from the split method and bootstrap method as a percentage of the mean estimate.
    \Explainbrackets{}
}%
\label{tab:split}
\end{table}

\section{Termination conditions}\label{sec:termination}                                                                                                                                                                        
The sensitivity to termination conditions can be far higher for parameter estimation than for evidence calculation.
This is both because parameter estimation can have much smaller sampling errors, and because the region close to the likelihood peak can have very high weight for some $f(\theta)$.
For example for the Gaussian likelihood~\eqref{equ:gaussian} an estimator such as $f(\theta)=\overbar{{|\theta|}^{-1}}$ may show significant errors due to termination conditions which were perfectly adequate for calculating $\logZ$.
Numerical tests in this paper use the termination conditions described by \citet[][Section 3.4]{Handley2015b}, stopping when the estimated evidence contained in the live points is less than $10^{-4}$ times the evidence contained in dead points.

When splitting runs into their constituent threads (\Cref{sec:divide}) then even in perfect nested sampling termination conditions must be chosen carefully to avoid causing differences between threads from different runs which terminate at different likelihoods.
This typically happens when
\begin{enumerate}
    \item termination conditions are worked out from the current set of dead points --- e.g.\ estimating the evidence $\Z$ remaining as in \citet[][Section 3.4]{Handley2015b}. This means some runs continue for longer than others;
    \item the final point which violates the condition is kept. This means threads from small runs are much more likely to have final points far exceeding the termination condition than threads from large runs.
\end{enumerate}
When comparing threads from different nested sampling runs, their equivalence can be maintained by using a termination condition which does not infer anything from the previous points, such as setting a fixed likelihood value $\like_\mathrm{term}$ for termination and discarding any point that exceeds it.
As we do not mix threads from different runs in our numerical tests we do not need this approach.

\section{Additional numerical tests: 3-dimensional Cauchy likelihood}\label{sec:additional_tests}

Table~\ref{tab:bootstrap_cauchy} shows numerical tests of sampling error estimates with a 3-dimensional Cauchy likelihood~\eqref{equ:cauchy} with a Gaussian prior~\eqref{equ:gaussian_prior}.
As in the 3-dimensional Gaussian case shown in~\Cref{tab:bootstrap}, the mean estimates of sampling errors from our bootstrap method are very close to measurements of sampling errors from repeated calculations --- this can be seen in the second row of Table~\ref{tab:bootstrap_cauchy}.
Again the empirical coverage rates for bootstrap credible intervals are close to their nominal values, as shown in the final two rows.

\begin{table}
\begin{tabular}{l c c c}
    \toprule
    & $\po$ & $\vart$ & $\poconf{84}$ \\
    \midrule
    Repeated runs \std{}                          & $ 0.044(0.3)               $ & $ 0.573(4)                 $ & $ 0.119(0.8)               $ \\
    Bootstrap \std{} / Repeats \std{}             & $ 1.005(7)                 $ & $ 1.003(8)                 $ & $ 1.002(8)                 $ \\
    Simulated $w_i$ $\std$ / Repeats $\std$       & $ 0.717(5)                 $ & $ 0.994(8)                 $ & $ 0.926(7)                 $ \\
    Bootstrap $\std$ estimate variation           & $ 9.3(1)\%                 $ & $ 12.7(2)\%                $ & $ 16.9(3)\%                $ \\
    Simulated $w_i$ estimate variation            & $ 8.0(1)\%                 $ & $ 12.0(2)\%                $ & $ 17.4(3)\%                $ \\
    Bootstrap $\mathrm{C.I.}_{95\%}$              & $ 0.072(5)                 $ & $ 6.70(7)                  $ & $ 1.69(2)                  $ \\
    Bootstrap Mean$\pm1\std$ coverage             & $ 68.6\%                   $ & $ 68.8\%                   $ & $ 68.7\%                   $ \\
    Bootstrap $\mathrm{C.I.}_{95\%}$ coverage     & $ 95.1\%                   $ & $ 92.1\%                   $ & $ 92.1\%                   $ \\
    \bottomrule
\end{tabular}
    \caption{Sampling errors for a 3-dimensional Cauchy likelihood~\eqref{equ:cauchy}, a Gaussian prior~\eqref{equ:gaussian_prior} and $n=200$.
The first row shows the standard deviation of 10,000 nested sampling calculations.
The second and third rows show the mean of 2,000 error estimates from the bootstrap and simulated weights methods respectively as a ratio to the error observed from repeated calculations; $200$ weight simulations and $200$ bootstrap replications were used for each run.
The fourth and fifth rows show the standard deviations of sampling error estimates for both methods as a percentage of the mean estimate.
The sixth row shows the mean of 100 bootstrap estimates of the one-tailed $95\%$ credible interval on the calculation result given the sampling error, each using 1,000 bootstrap replications.
The final two rows show the empirical coverage of the bootstrap standard error and $95\%$ credible interval from the 10,000 repeated calculations.
\Explainbrackets{}
}%
\label{tab:bootstrap_cauchy}
\end{table}

\end{appendices}

\end{document}